\documentstyle[]{mn}
\input epsf
 
\def\BE{\begin{equation}}
\def\EE{\end{equation}}
 
\def\BA{\begin{eqnarray}}
\def\EA{\end{eqnarray}}
 
\newcommand{\rmn}[1] {{\rm #1}} 
\title[Chemical evolution of gas-rich dwarf galaxies]{The chemical evolution of
 gas-rich dwarf galaxies}

\author[T.I. Larsen, J. Sommer-Larsen \& B.E.J. Pagel]{T.I. Larsen$^{1}$,
J. Sommer-Larsen$^{2}$, B.E.J. Pagel$^{3}$ \\
$^{1}$Copenhagen University Observatory, Juliane Maries Vej 30, DK-2100
Copenhagen {\O}, Denmark 
(e-mail: thommy\underline{ }l@astro.ku.dk) \\
$^{2}$Theoretical Astrophysics Center, Juliane Maries Vej 30, DK-2100 Copenhagen
{\O}, Denmark 
(e-mail: jslarsen@tac.dk) \\ 
$^{3}$Astronomy Centre, CPES, University of Sussex, Brighton BN1 9QJ, UK 
(e-mail: bejp@star.cpes.susx.ac.uk)}

\date{Received.........; in original form..........}

\begin{document}

\maketitle

\begin{abstract}
A numerical double burst model of  the chemical evolution of gas-rich dwarf galaxies 
has been developed. The model is fitted to a sample of N/O, O/H, Y and gas 
fraction observations, where N/O and O/H are the relative abundances by number of 
nitrogen to oxygen and oxygen to hydrogen, respectively. Y is the abundance by 
mass of helium. Closed models as well as models including enriched outflow, 
ordinary outflow and ordinary outflow combined with inflow are considered.The 
bursts are assumed to be instantaneous but ordered in pairs to explain
the scatter in N/O-O/H. The method of gas fraction fitting is revised, and it 
is found that it is very important to specify whether dwarf irregulars (dIrrs) or 
blue compact galaxies (BCGs) are considered. Effective enriched winds fail when 
fitting N/O, whereas closed models, models with ordinary winds or a combination 
of ordinary winds and inflow are all viable.
\end{abstract}
\begin{keywords}
methods:numerical -- galaxies:irregular -- galaxies:starburst -- 
galaxies:abundances -- ISM:\hbox{H\,{\sc ii}}-regions
\end{keywords}

\section{Introduction}
The chemical evolution of dwarf irregular (dIrr) 
\footnote{In this paper a dwarf galaxy is defined to have absolute magnitude 
M$_{\rm B}\geq -17$.} and blue compact emission-line 
galaxies (BCGs) is of particular interest  because a substantial body of 
observational data is available and some degree of simplicity exists because 
of the low level of `metal' enrichment and absence of large abundance gradients.  
Furthermore, their wide range of intrinsic properties makes them suitable 
objects 
for testing certain expectations from stellar nucleosynthesis theory and the 
`Simple' or other models of galactic chemical evolution, although at the same time 
there are complications associated with inflow of unprocessed material, outflow 
in homogeneous or selective galactic winds and bursting (or `gasping') modes of 
star formation. Chemical evolution models attempt to apply all these concepts to 
account for the distribution of different elements, notably helium, oxygen and 
nitrogen, in relation to star formation rates and gas fractions. Because many 
parameters such as these last two are generally very poorly determined, the most 
convincing tests come from the comparison of different elements with one another. 

Back in the 1970s, Smith \shortcite{smith:1975}, Peimbert \shortcite{peimbert:1978}
and Edmunds \& Pagel \shortcite{edmunds:1978} 
noticed a contribution of primary nitrogen to the N/O ratio in Galactic and 
extragalactic 
H II regions with low oxygen abundance and Edmunds \& Pagel attributed the 
existence of scatter in N/O at a given O/H to the existence of a time delay 
in primary nitrogen production by intermediate-mass stars, combined with 
differing effective ages of the underlying stellar populations.  Alloin et al. 
\shortcite{alloin:1979} also noted the primary nitrogen and attributed scatter in N/O
to variations 
in the initial mass function (IMF), whereas Lequeux et al. \shortcite{leq:1979} in 
their classic study of helium, nitrogen and oxygen in irregular galaxies and BCGs 
confirmed the primary nitrogen likewise, but were 
not convinced that there was any real scatter in their data. The models of 
Alloin et al. and Lequeux et al. assumed evolution to take place smoothly as a 
function of time;  Matteucci \& Chiosi \shortcite{mc:1983} were the first to incorporate 
into chemical evolution models for these systems the idea of bursting modes of 
star formation  as prevously inferred by Searle \& Sargent \shortcite{searle:1972} and 
Searle, Sargent \& Bagnuolo \shortcite{bagnuolo:1973}, and interpreted on the  
basis of the SSPSF hypothesis by Gerola, Seiden \& Schulman \shortcite{gerola:1980}.  
The basic pattern 
of a primary (constant N/O) pattern at low metallicities in H~II regions 
changing over to a secondary pattern (N/O $\propto$ O/H) at higher ones has 
been confirmed in many more recent investigations (e.g. Vila-Costas \& Edmunds 
1993; van Zee, Salzer \& Haynes 1998).  
   
The chemical evolution of dIrrs and BCGs has been studied in many more recent 
investigations.  Matteucci \& Tosi \shortcite{mt:1985} found good fits to the data with 
a Salpeter IMF, inflow, homogeneous outflow, bursting star formation 
 and a choice of third dredge-up 
parameters from Renzini \& Voli \shortcite{RV:1981}, attributing scatter in the N/O ratio 
to variations in $M_{\rm up}$, the upper limit to the masses of stars 
undergoing the third dredge-up with hot-bottom burning. Garnett \shortcite{garnett:1990} 
indicated schematically how the occurrence of bursts could in itself lead to 
variations in the N/O ratio just as a result of observing systems at different 
stages in the burst cycle.  Pilyugin \shortcite{pilyugin:1992,pilyugin:1993} developed 
similar ideas in 
quantitative numerical models involving self-enrichment of H II regions 
and selective galactic winds as well as bursting modes of star formation, and 
Marconi, Matteucci \& Tosi \shortcite{marconi:1994} also developed models with 
bursts and selective winds, while Carigi, Col\' in \& Peimbert \shortcite{carigi:1998}
have investigated 
similar models, but prefer a `bottom-heavy' IMF similar to one claimed in some 
globular clusters and giving rise to low true yields.  However, part of 
the motivation for invoking selective winds 
was the apparent existence of a large $dY/dZ$ ratio suggested by Pagel et al. 
\shortcite{pagel:1992}, which no longer seems valid \cite{izotov:1998}, and the scatter
in N/O also seems to have been overestimated in those investigations. In this paper,
therefore, we investigate the problem again, making use of more recent 
data and models of stellar nucleosynthesis and exploring in particular the role 
of mixing processes and of differing burst phases in leading to scatter in the 
N/O, O/H relation and the relationship between oxygen abundance and gas fraction.

The structure of the article is as follows: Section\ \ref{abundance} presents
the adopted sample of abundance observations. Section\ \ref{mix} discusses the
evolution of \hbox{H\,{\sc ii}}-regions, wind-driven bubbles and
supernova-driven supershells to investigate possible mixing scenarios.
This leads to the description of our double-bursting models in section\ 
\ref{model}. The results of fitting the models to the sample of observations are
presented in section\ \ref{result}, and discussed in section\ \ref{conc}. 
\section{The observational data} \label{abundance}
It has been stressed that the sample should consist mainly of BCGs, as these 
objects are extreme in both star formation rates
and metallicity, thus being attractive for a bursting model. Another
demand has been to include the most recent data only, in order to get the
observations as accurate as possible. Further, a few damped Ly-alpha systems 
(DLA) have been included, mainly for the sake of comparison. The motivation for
including them is found in the growing opinion that they represent dwarf 
galaxies in
early (high z) stages of evolution. Thus, they may offer some insight into the
extreme low-metallicity environment in the early evolution of present-day 
dwarf galaxies.

The following sample is selected: BCGs
and dIrrs from Pagel et al. \shortcite{pagel:1992}, BCGs from Izotov, Thuan
\& Lipovetsky 
\shortcite{izotov:1997}, the BCG SBS 0335-052 from Izotov et al.
\shortcite{izotov:1997b}, the BCG IZw18 from Izotov \& Thuan 
\shortcite{izotov:1998},
and finally four DLAs from Lu, Sargent \& Barlow \shortcite{lu:1998}.
From the sample of Pagel et al. \shortcite{pagel:1992} are excluded LMC, 
NGC5253
and NGC5455/NGC5461 (the last two are \hbox{H\,{\sc ii}}-regions in the nearby 
spiral M101),
and objects included from one of the other sources.
\begin{figure}
\centering
\epsfxsize=6cm
\epsfysize=4cm
\epsfbox{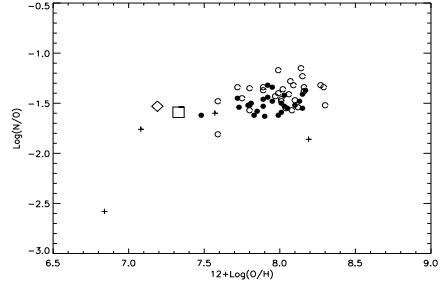}
\caption[]{The observational sample used in this paper. The
symbols are: diamond=IZw18 \cite{izotov:1998}, square=SBS0335-052
\cite{izotov:1997b}, filled circles are BCGs \cite{izotov:1997} and open
circles are BCGs and dIrrs \cite{pagel:1992}. Finally, the plusses are
DLA-systems \cite{lu:1998}.\label{sampleNO}}
\end{figure}
\begin{figure}
\centering
\epsfxsize=6cm
\epsfysize=4cm
\epsfbox{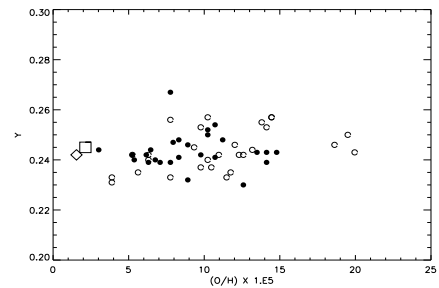}
\caption[]{The helium abundance sample. The symbols
are as in fig.\ \ref{sampleNO} No DLAs are shown.
 \label{sampleY}}
\end{figure}
The total sample is shown in fig.\ \ref{sampleNO} for N/O-O/H and 
fig.\ \ref{sampleY} for Y-O/H.
The N/O abundances seem to have a constant level for 12+log(O/H) less than 8,
implying the major part of the nitrogen to be produced as a primary element.
For 12+log(O/H) higher than about 8, N/O is increasing as a function of 
metallicity, suggesting the major part of the nitrogen to be secondary.
These trends are wellknown, see e.g. Garnett \shortcite{garnett:1990}, 
Vila-Costas \& Edmunds \shortcite{vila:1993}, Pettini, Lipman \& Hunstead
\shortcite{pettini:1995} and van Zee, Salzer \& Haynes \shortcite{vanzee:1998a}.

Further, the scatter in N/O seems to be significant and our models are based 
on this assumption. If we fit a straight line to our selected observations, 
ignoring the DLAs, we find the standard deviation in N/O to be $\sigma $=0.11,
which is of the same order of magnitude as the observational uncertainty 
\cite{izotov:1997}, but a real scatter is evident in other data from a wider
range of sources, see e.g. Kobulnicky \& Skillman \shortcite{kobul:1996}.

The DLA-systems pose severe problems when observing N and O abundances. Usually
the N-lines are occuring on top of underlying absorption, and O-lines are 
almost always saturated. Because of these difficulties, Lu et al. 
\cite{lu:1998} found it useful to use Si or S
instead of O, which makes sense as O/Si and O/S are found to be the same as
solar in both the Galactic disc and halo as well as in 
\hbox{H\,{\sc ii}}-regions in nearby galaxies. The question is whether the 
abundance ratios are equal to solar at the extreme low metallicity of DLAs. 
This is by now the largest
uncertainty in this method. Thus, in fig.\ \ref{sampleNO}, the DLAs are
represented by their N/Si and Si/H, assuming that these ratios are
equivalent to N/O and O/H. Four systems have been selected from the Lu et al. 
sample namely those in front of QSO's 0100+1300, 1331+1704, 1946+7658 and 
2343+1232. For the rest of their sample, the quality of the observations 
restricts the abundance determination to be performed as higher or lower limits. 
Lu et al. did not make 
any corrections for dust. However, they noted that the depletion in dust is 
less than 0.4 dex.
Though the observations of the four DLAs are included, they should not be 
taken too seriously and are not given much attention in this paper.
 
The helium mass fraction shows a linear dependency on metallicity.
This linearity has been used for extrapolation back to O/H=0, giving the
primordial He abundance. Izotov et al. \shortcite{izotov:1997} used their
observations, included in the present sample as well (filled circles in fig.\
\ref{sampleY}), to obtain 0.243$\pm $0.003, using a linear fit with slope
dY/dZ=1.7$\pm $0.9 and assuming Z=20(O/H). Error bars are omitted in the 
figure to keep the clearness. However, the error bars are included in the 
figures presenting our results in section \ref{result} and show no 
observational evidence for a scatter.
Omitting SBS 0749+568, represented by the isolated point above 0.26 in fig.\
\ref{sampleY}, but including
all other galaxies in the sample, a linear least-square fit gives
dY/dZ=2.63$\pm $2.21, again using Z=20\ (O/H), and assuming a confidence
interval of 95 per cent. This gives a primordial helium abundance 
$Y_p$=0.238$\pm $ 0.004, which is consistent with Izotov et al. 
\shortcite{izotov:1997} within uncertainties.

\section{Ejecta dispersal and mixing} \label{mix}
Many chemical evolution models assume a one-zone description with 
instantaneous mixing. Assigning 
the term one-zone to a BCG seems to be quite a poor approximation, and assuming 
the mixing to be instantaneous will always
be doubtfull. However, if we choose to reject the assumptions, we are faced
with the problem that no complete theory of mixing is available at present.
The problem is that the gas dynamics following a starburst are very complex. The
energy input into the ISM comes from photoionization, stellar winds from
massive stars followed by SN type II.
Different scenarios have been proposed for the dispersal and mixing processes.
One of the propositions has been 'self-enrichment', referring to the suggestion 
that only the \hbox{H\,{\sc ii}}-region, surrounding the newly formed stellar 
cluster, is enriched with heavier elements, in particular oxygen. Thus, the 
observed abundances, using emission lines,
may be considerably higher than they would be if observed in the neutral 
medium. The idea was originally proposed by Kunth \& Sargent 
\shortcite{KS:1986}, suggesting that the enrichment is confined to take 
place within the Str\"{o}mgren-sphere. However, the 
hypothesis is controversial, and a lot of opposing arguments exist. Use of 
self-enrichment implies the assumption of almost instantaneous
mixing within the \hbox{H\,{\sc ii}}-regions. The problem is that the
correlated energy input of SNe changes the physical conditions of \hbox{H\,{\sc
ii}}-regions dramatically. In fact, the density is decreased by a factor $ 10^3
- 10^6$ and the temperature is increased by a similar factor behind the 
SN-driven shock front. The observed lines of single or
double ionized N and O cannot arise from such extreme conditions. A future
project could be X-ray abundance observations. If the observed 
emission lines cannot form within the superbubble/supershell (hereinafter
a 'superbubble' is a wind-driven bubble, while a 'supershell' is a SN-driven
shell), they may arise from a region outside the shock front. But according to 
detailed numerical hydrodynamical models
\cite{Tenorio-Tagle:1996}, the ejecta will stay within the
superbubble/supershell for a large part of their evolution, not enriching the
surrounding medium. To test the viability of this statement, a few calculations 
are presented below, comparing radii of superbubbles/supershells and radii of 
\hbox{H\,{\sc ii}}-regions.
Further, observations give evidence against the 
self-enrichment hypothesis, see e.g. van Zee et al. \shortcite{vanzee:1998b},
Kobulnicky \shortcite{kobulnicky:1997}, Kobulnicky \& Skillman 
\shortcite{kobulskill:1997} and Kobulnicky \& Skillman 
\shortcite{kobulnicky:1998}.

\subsection{H\,{\sevensize\bf II}-region evolution}
\begin{figure}
\centering
\epsfxsize=\linewidth
\epsfbox{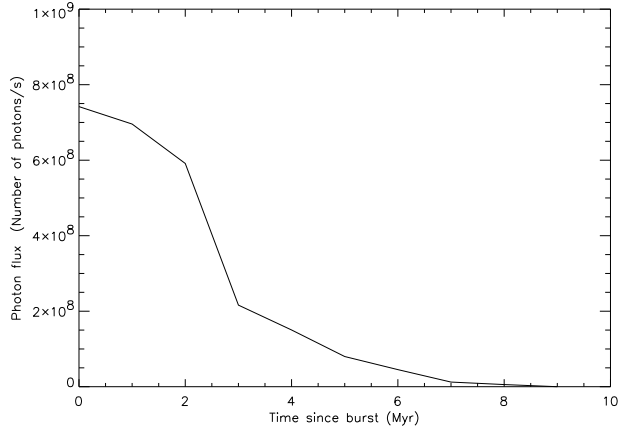}
\caption[]{The decrease in ionizing photon flux counted from a starburst 
event.} 
\label{photdec}
\end{figure}

In the following, the radii of \hbox{H\,{\sc ii}}-regions are calculated for 
comparison with superbubble/supershell radii. The Str\"{o}mgren-sphere is 
defined to be the sphere within which all ionizing photons are absorbed. Thus,
setting the flux of ionizing photons equal to the number of recombinations,
integrated over the entire Str\"{o}mgren-sphere gives:
 \BA
 \frac{4\pi }{3}R(t)^3\alpha _{B}n_en_p\epsilon =N(t)   \label{sphere}
 \EA
\noindent 
$\epsilon $ is the filling factor, $\alpha _B$ is the recombination rate
equal to $2.59\times 10^{-13} \rmn{cm}^3\rmn{s}^{-1}$
(T=10000K) \cite{osterbrock:1989}.
$n_p$ and $n_e$ are the number densities of protons and electrons,
respectively, and $N(t)$ is the flux of ionizing $Ly_c$-photons.
Assuming that all hydrogen inside of the Str\"{o}mgren-sphere is fully
ionized, $n_e \sim n_p$. 
The filling factor is an indicator of the uniformity of matter, having 
a value of 1 when the matter is completely uniformly distributed. A typical
value of the filling factor is of the order 0.01 \cite{kennicutt:1984}. The
electron density in \hbox{H\,{\sc ii}}-regions is typically $n_e \sim \ 100\ 
\rmn{cm}^{-3}$ \cite{izotov:1997}. When inserting these two values into eq.\ 
\ref{sphere}, the corresponding
rms density is $n_e = 10\, \rmn{cm}^{-3}$, since $\epsilon = 1$
and $n_e = 10\, \rmn{cm}^{-3}$ is equivalent to the insertion of the observed 
values.  
According to Spitzer \shortcite[p.251]{spitzer:1978}, the first phase in the 
existence of an \hbox{H\,{\sc ii}}-region is characterized by an almost static 
Str\"{o}mgren-sphere. Henceforth, the radius of this initial sphere will be 
referred to as 'the initial Str\"{o}mgren radius', $R_0$. During the second
phase, the sphere is expanding until the \hbox{H\,{\sc ii}}-region vanishes.
If the flux of ionizing photons is assumed constant for the moment, 
Spitzer \shortcite{spitzer:1978} found the relation
 \BA
 R(t)=R_0\left( 1+{\frac{7}{4}}\frac{C_{II}t}{R_0}\right) ^{\frac{4}{7}}
 \label{Spitzer}
 \EA
\noindent on the basis that the expansion velocity nearly equates the velocity of 
the associated shock. $t$ is the time since the region 
started to expand, virtually equal to the age of the burst, and $C_{II}$ 
is the sound speed in the \hbox{H\,{\sc ii}}-region, equal to $17\rmn{km}\, 
\rmn{s}^{-1}$ (T=10000K, $\gamma =5/3$) \cite{osterbrock:1989}.
Eq.\ \ref{Spitzer} is not 
realistic, though, because the photon flux is decreasing in time. In the 
following this is taken into account, extending the calculations by Spitzer.
Differentiating eq.\ \ref{sphere} with respect to time gives
\BA \label{difN}
\frac{dN}{dt}=3\beta R^2\rho _{II}^2\frac{dR}{dt}+2\beta 
R^3\rho _{II}\frac{d\rho _{II}}{dt} 
\EA
\noindent where $\beta =\frac{4\pi }{3}\eta ^2\alpha _B\epsilon $, $n_e=\eta 
\rho _{II}$, so $\eta $ gives the relation between the number density of 
electrons and the mass density. Below it is shown that $\eta $ cancels in the 
calculations, so we will not worry about its value. Finally
$\rho _{II}$ is the mass density within the \hbox{H\,{\sc ii}}-region. 
Consider a shell of ionized gas. Assuming uniform expansion, i.e.
\BA 
\frac{1}{r}\frac{dr}{dt}=\frac{v_i}{R}
\EA
\noindent where $r$ is the comoving radius of the shell and $v_i$ is the 
velocity of the ionized gas just within the ionization front, assumed to
have the same radius $R$ as the Str\"{o}mgren-sphere. The mass inside the 
comoving shell is conserved by definition, so $r^3\rho _{II}$ is constant 
leading to
\BA
\frac{1}{r}\frac{dr}{dt}=-\frac{1}{3\rho _{II}}\frac{d\rho _{II}}{dt}
\EA
Inserting these two equations into eq.\ \ref{difN} gives
\BA \label{intermed}
v_i=\frac{1}{2}\frac{dR}{dt}-\frac{\frac{dN}{dt}}{6\beta R^2\rho _{II}^2}
\EA
\noindent and $\frac{dR}{dt}=V_i$, the velocity of the ionization front. 
The physical conditions across the shock, formed ahead of the ionization front
 may be described by the jump condition
\BA \label{jump}
\rho _IV_{s}^2=p_{II}+\rho _{II}u_s^2
\EA
\noindent $\rho _I$ being the mass density of the surrounding neutral medium, 
$V_s$ is the velocity of the shock, $p_{II}$ is the pressure within the 
\hbox{H\,{\sc ii}}-region and $u_s$ is the inward velocity of matter with 
respect to the
shock. It has been assumed that the density and pressure between the shock
and ionization fronts are the same as within the ionization front, and the 
pressure of the surrounding neutral medium has been neglected. Assume 
$V_s=V_i$, so $u_s=u_i=V_i-v_i$, where $u_i$ is the inward velocity of matter 
with respect to the ionization front. Inserting this and the 
equation of state $p_{II}=\frac{\rho_ {II}C_{II}^2}{\gamma}$ with $\gamma 
=5/3$, $\rho_ {II}=\sqrt{\frac{N}{\beta R^3}}$ (from eq.\ \ref{sphere}) and
$\rho _I=\sqrt{\frac{N_0}{\beta R_0^3}}$ (i.e. the density before
expansion starts) into eq.\ \ref{jump} and\ \ref{intermed}, one finally arrives
at the second order equation
\BA
\left[ \left( \frac{NR_0^3}{N_0R^3}\right) ^{-1/2}-\frac{1}{4}\right] \left(
\frac{dR}{dt}\right) ^2-\frac{\frac{dN}{dt}}{6\beta R^2\rho _{II}^2} \left(
\frac{dR}{dt}\right) \\ \nonumber
- \left( \frac{\frac{dN}{dt}}{6\beta R^2\rho _{II}^2}
+\frac{C_{II}^2}{\gamma}\right) =0
\EA
\noindent It is straightforward to solve this equation for $\frac{dR}{dt}$, 
obtaining a differential equation, which is solved numerically, using
timesteps equal to 1 Myr. For the calculations we used $n_e=10\, 
\rmn{cm}^{-3}$ ($\beta =1$)as the initial density. When inserted into the 
second order equation instead of $\rho _{II}$, the
factor $\eta $ cancels remembering that it is included in $\beta $ as well.
For every timestep, the density $n_e(t)$ is calculated from eq.\ \ref{sphere}.
The fluxes of ionizing photons are taken from the models by Stasinska \& 
Leitherer \shortcite{leitherer:1996}, giving the fluxes from 1 Myr to 10 Myr
after 
the burst in intervals of 1 Myr. An instantaneous starburst of mass $10^6 
M_{\odot }$ was assumed, comparable to those used in our model. Further, a 
Salpeter IMF with the same upper mass limit is used, only demanding a simple 
scaling from their lower mass $1 M_{\odot }$ to our lower mass, either $0.1$ 
or $0.01 M_{\odot }$. 
If $m_L=0.1$ the normalization constant is 0.17, see section \ref{imf}. 
Integration of the IMF yields a mass fraction of 0.39 above 1 $M_{\odot }$, so 
the adopted fluxes are multiplied by 0.39. 
Correspondingly, using $m_L=0.01$ the normalization constant is 0.07, 
giving a mass fraction of 0.17 above 1 $M_{\odot }$. The decrease of continuum
photon flux calculated by Stasinska \& Leitherer using their parameter values 
is shown in figure\ \ref{photdec}.

$R_0$ is found from eq.\ \ref{sphere} by inserting the initial flux of photons
and the initial density $n_e=10\, \rmn{cm}^{-3}$. The initial flux, 
$N_0=N(t=0)$ is the flux just before the \hbox{H\,{\sc ii}}-region starts to 
expand. No information exist on the photon flux before 1 Myr after the burst. 
Thus, $N_0=N(1\rmn{Myr})$ is assumed so $R_0$ is equal to R(1 Myr) found from 
eq.\ \ref{sphere}. The calculations are performed for both of the IMF's and 
the results are shown in table\ \ref{radii}. 

\subsection{Supershell evolution} \label{super}
In the case of a single supernova event, it is rather simple to follow the four
main evolutionary phases of its remnant analytically. First, the ejecta and the 
leading shock move almost undecelerated. Second, the velocity of the shock 
decreases, as a shell of swept-up ISM builds up and grows larger.  In this 
phase, radiation from the hot cavity is not important, and the total energy 
within the shock is conserved. Third, radiative cooling of all matter within 
the shock become important. The shell is now driven by the pressure still existing 
in the cavity. Finally the PdV work from the cavity vanishes, so that the shell 
is now only being driven by the momentum. At some moment, the velocity of the 
shell has dropped to a value comparable to the rms velocity of the ISM, thus 
being non-detectable.
 
When one deals 
with the correlated effect of many supernovae, which is the case in starburst
galaxies where the number of massive stars is large, additional considerations
have to be made, since supernovae keep feeding energy to the cavity for about 
50 Myr. The idea with the following calculations is to get a feeling
of the timescales and radii of supershells. The analytical expressions from 
McCray \& Kafatos \shortcite{McCray:1987} are adopted to do the simple 
calculations, and the equations below are taken from their article, unless 
another reference is given. Strictly
speaking, the analytical expressions may be used only if the ambient density
is constant. However, the supershell first expands into the wind-driven bubble 
environment, then into a region with density depending on the \hbox{H\,{\sc 
ii}}-region expansion, and finally into a non-expanding region outside the
\hbox{H\,{\sc ii}}-region. Hence, the evolution of the supershell is only
followed while it is smaller than the \hbox{H\,{\sc ii}}-region, and the 
density within the superbubble is assumed to be the same as outside. This last 
assumption is not realistic, as the passage of the wind-blown shock sweeps
a large fraction of the gas into a superbubble. Hence, we may expect a more 
accurate model to show that the supershell reaches the superbubble at an 
earlier time.
The gas gathered by the wind-driven shock cools to form the superbubble 
after relatively short time

\BA \label{bcool}
t_b=(2.3\times 10^4 \rmn{yr})n_e^{-0.71}L_{38}^{0.29}
\EA
\cite{Maclow:1988}, and the radius of the superbubble is 
\BA \label{bubble}
R_s = 269\ pc\ (L_{38}/n_{e})^{1/5}t_{7}^{3/5}
\EA
 \noindent where $L_{38}$ is the characteristic wind luminosity in units of 
$10^{38}\rmn{ergs}\, \rmn{s}^{-1}$ and $t_7$ is the elapsed time since the 
starburst in units of $10^7$yr. Strictly speaking, the time is counted from 
the start of superbubble 
expansion, but since O-stars start their wind-phase soon after the burst, 
$t_7$ is counted from the burst. $n_{e}$ is the matter density within the 
\hbox{H\,{\sc ii}}-region, assumed to be homogeneous.
The wind luminosity is found in Leitherer \& Heckman 
\shortcite{leitherer:1995} from their fig. 55 (instantaneous starburst, 
metallicity is one tenth of solar, mass of burst is
$10^6 M_{\odot }$, Salpeter IMF with upper limit 100 $M_{\odot }$ and lower
limit 1$ M_{\odot }$). At $10^6$ yr the wind luminosity is read-off to be
$10^{39}\, \rmn{ergs}\, \rmn{s}^{-1}$. Doing the same IMF scaling as in the 
photon flux
discussion above, $L_{38}=3.9$ is obtained if $m_L=0.1 M_{\odot }$ and 
$L_{38}=1.7$ if $m_L=0.01 M_{\odot }$. 
The wind phase is important for as long as O-stars exist ($\approx 5$ Myr).
However, after about 3 Myr the first supernovae explode and soon dominate
the total energy output. The superbubble radius has been calculated 1, 2 and 
3 Myr after the burst using eq.\ \ref{bubble}. The density typical for an 
\hbox{H\,{\sc ii}}-region during its early evolution is $10\, \rmn{cm}^{-3}$, 
as discussed above. 

After 3 Myr, the first SN appear. Hence, the expansion of the supershell is set
to start at t=3 Myr. Strictly speaking, the energy output from supernovae 
appears as discrete events. However, it can be treated as continuous, as long 
as the interval between explosions is sufficiently short, at most in the order 
of $10^5$ yr (Tomisaka, Habe \& Ikeuchi\ 1981).
If the total mass of the burst is $10^6 M_{\odot }$, the number of stars with 
mass greater than or equal to 8$M_{\odot }$ (lower limit of a SNII progenitor,
according to Woosley \& Weaver \shortcite{woosley:1986}) may be calculated 
using our IMF. The result using the 0.1 $M_{\odot }$ lower mass limit is 8145 
stars, whereas it is 3452 stars using $m_L=0.01 M_{\odot }$. The event 
lasts for about $5\times 10^7$yr so the mean interval between explosions is in 
the order of $6\times 10^3$ yr and $1\times 10^4$ yr, respectively. 
Thus, it is meaningful to assume continuous energy injection.
Using this, eq.\ \ref{bubble} may be used with just one
replacement, namely the insertion of the mean supernova power instead of the
wind luminosity. The mean power is calculated by dividing the total SN energy
$N_{\star }E_{51}\times 10^{51}$ ergs with the total duration of the event.
$N_{\star }$ is the number of stars with mass greater than 8 $M_{\odot }$
and $E_{51}$ is the energy in units of $10^{51}$ ergs. Here, 0.4 is adopted as 
a representative value, see Woosley \& Weaver 
\shortcite[their table 1]{woosley:1986}. The energy ejection is assumed not 
just to be continuous, but also constant in time. This is probably quite a good 
approximation, see Leitherer \& Heckman (1995). The considerations lead to
 
\BA  \label{adiabat}
R_s &=& 97\ \rmn{pc}\ (N_{\star }E_{51}/n_{e})^{1/5}t_{7}^{3/5} \\ \nonumber
V_s &=& 5.7\ \rmn{km}\ \rmn{s}^{-1}(N_{\star }E_{51}/n_{e})^{1/5}t_{7}^{-2/5}
\EA
\noindent
The time of shell formation is given by an expression similar to eq.\ 
\ref{bcool}, obtained by replacing the wind luminosity with the mean supernova
power.

The density of the \hbox{H\,{\sc ii}}-region is assumed constant in time and 
equal to 3\ $\rmn{cm}^{-3}$, a reasonable value during late stage 
\hbox{H\,{\sc ii}}-region evolution (see table\ \ref{radii}).
Eqs.\ \ref{adiabat} are used in the adiabatic phase only (no cooling), since they
implicitly assume that the loss of energy within the shell is negligible.
Cooling, and hence energy loss, from the interior becomes important at a 
time
\BA \label{timescale}
t_c = 4\times 10^6\ \rmn{yr}\ \xi ^{-1.5}(N_{\star }E_{51})^{0.3}n_{e}^{-0.7}
\EA
\noindent 
At this time, the shell has a radius
\BA
R_c = 50\ \rmn{pc}\ \xi ^{-0.9}(N_{\star }E_{51})^{0.4}n_{e}^{-0.6}
\EA

\noindent found by inserting eq.\ \ref{timescale} into eq.\ \ref{adiabat}. 
$\xi $ is the metallicity in units of the solar metallicity. For the objects
in question, $\xi =0.1$ is a representative value. Even if a small starburst 
mass of 10 $M_{\odot }$ is assumed and the density of the surrounding medium 
is 3\ $\rmn{cm}^{-3}$, it turns out that the supershell reaches the end of the
adiabatic phase at a time $2.1\times 10^7$ yr if $m_L$=0.1 is used and 
$1.6\times 10^7$ yr if $m_L$=0.01. Hence, the interior of the supershell 
starts to cool after the \hbox{H\,{\sc ii}}-region has vanished. This 
tells us that we may use eqs.\ \ref{adiabat} to 
calculate the size of the supershell, as we are only interested in the 
evolution during \hbox{H\,{\sc ii}}-region existence.
The results are given in table\ \ref{radii}.

\begin{table}
\caption{\hbox{H\,{\sc ii}}-regions and supershells} \label{radii}
\begin{tabular}{cccc}
 $t$(Myr) & $R_{\hbox{H\,{\sc ii}}}$ (pc) & $R_{s}$ (pc) & 
$n_e$ ($\rmn{cm}^{-3}$) \\
 1 & 209 (158)& {\bf{56}} ({\bf{47}})& 10.0 (10.0) \\
 2 & 222 (172)& {\bf{85}} ({\bf{72}})&  8.8 (8.5) \\
 3 & 233 (184)& {\bf{108}} ({\bf{92}})&  7.5 (7.1) \\
 4 & 239 (189)& 123 (103)&  4.4 (4.1) \\
 5 & 245 (196)& 186 (157)&  3.5 (3.3) \\
 6 & 250 (201)& 237 (200)&  2.5 (2.3) \\
 7 & 255 (206)& 282 (238)&  1.8 (1.7) \\
\end{tabular}

\medskip The radii and densities of \hbox{H\,{\sc ii}}-regions compared 
to the radii of superbubbles/supershells at 
different times after the burst calculated with the $m_L$=0.1$M_{\odot}$ 
cutoff. Values of shell radii typed in bold, are the 
radii of the superbubbles. For t$>$3 Myr, the calculated shell 
radii are for the supershells. The values for $m_L$=0.01$M_{\odot}$ are 
given in brackets. The size of the superbubble at 4 Myr after the burst is 
roughly the same as the size of the supershell, so the supershell reaches 
the superbubble about 1 Myr after the appearance of the first SNe.
\end{table} 

The measured emission lines may indeed originate in the part of the 
\hbox{H\,{\sc ii}}-region that is still outside the superbubble, since the size of the 
\hbox{H\,{\sc ii}}-region always exceeds the size of the superbubble as seen in 
table\ \ref{radii}. It also exceeds the size of the supershell until 6 Myr 
after the burst, though only significantly within the first 5 Myr. The above 
calculations do not prove that the emission lines arise outside the 
superbubble/supershell - they just confirm that it is a possibility. If 
emission lines really are measured this way, they are {\underline{not} 
affected by enrichment from the present burst, thus opposing self-enrichment.

\subsection{Propagating star formation}
Propagating star formation means that the energy deposited in the ISM by an 
evolving star formation event, is initiating new star formation. As a
supershell grows, it will become Rayleigh-Taylor unstable and fragments. The 
situation has been treated by Elmegreen \shortcite{elmegreen:1994}, who 
found an expression for the timescale of cloud-collapse in the supershell:

\BA 
t_{cloud}=103\, \left( \frac{n_0\mathcal{M}}{\rmn{cm}^{-3}}\right) ^{-1/2}\; 
\rmn{Myr}
\EA
where $\mathcal{M}$ is the expansion speed of the shell divided by the rms 
velocity dispersion in the shell. For an adiabatic shell, $\mathcal{M}$ is 
equal to 1.8
\cite{elmegreen:1994}. If $n_0=10\ \rmn{cm}^{-3}$, this gives the result that
star formation is expected to start not earlier than $\sim 24$ Myr after 
shell formation. On the other hand, if $n_0=3\ \rmn{cm}^{-3}$ as used in the 
calculations above for a typical density in 
the late stages of \hbox{H\,{\sc ii}}-region evolution, the timescale is more 
like 44 Myr. The mass of the supershell is given as
\BA \label{shell}
M=\rho _0\, V_{ss} \approx 1.3 n_0 m_p \frac{4\pi }{3} R^{3}
\EA

\noindent assuming all of the ISM originally within $R$ to be incorporated
in the shell. $\rho _0$ is the mass density of the ambient medium, $V_{ss}$ is 
the volume occupied by the hot phase within the supershell, $m_p$ is the 
proton mass and 1.3 is the approximate mass per particle in units of the proton 
mass. The radius is of the order a few
100 pc as shown above, so eq.\ \ref{shell} gives a shell mass of the order 
$10^6 M_{\odot }$. Hence the induced star formation 
is of the same order of magnitude as the original central burst.

Observational indications on the existence of star forming supershells are
numerous. One recent example is NGC 2537 included in the sample of Martin
\shortcite{martin:1998}. In $H_{\alpha }$, it shows a very clear spherical
distribution of starformation sites.
As a curiosity, it may be mentioned that Mori et al. \shortcite{mori:1997} 
calculated the effects of star formation in an expanding supershell, using
a 3-D hydrodynamical code including a dark matter halo. The results were in
remarkable accordance with available observations of dE's, such as exponential
surface brightness profile, positive metallicity gradient and inverse color
gradients. 

If the ejected metals from the central burst mix with the material of the
supershell, the \hbox{H\,{\sc ii}}-regions of the induced burst will show 
abundances differing from the central \hbox{H\,{\sc ii}}-region. Although there 
have been some doubt whether the metals are allowed to mix into the supershell
\cite{Tenorio-Tagle:1996}, 
and if so when this will happen, we have been inspired by this possibility to 
let our model have a double burst nature.

\section{The model} \label{model}
We have designed a model to fulfil certain demands, namely that
the observed trend of constant N/O for low metallicities and increasing
N/O for higher metallicities should be reproduced and also that 
the observed scatter in N/O should be explained.
Fixing all the parameters by fitting the N/O-O/H evolution, the model
should also be able to explain the observed helium mass fraction as a
function of O/H and O/H as a function of gas fraction.

The included elements are H,He,C,N,O. The production of N is still a
hot topic, since the degree of primary production at various metallicities
is unclear. Further, N is produced mainly by intermediate mass stars 
while O is produced by massive stars, making it necessary to include 
stellar lifetimes. However, the inclusion of both N and O provides  the
opportunity of constraining stellar parameters and mixing scenarios.

To account for the scatter in N/O, it is suggested that the bursts are
instantaneous and ordered in pairs, hence using the delay between the
ejection of N and O. The principle of time delay  was used by Garnett 
\shortcite{garnett:1990}, though employing single bursts only. It is 
doubtful whether single bursts produce scatter, since at least some abundance
observations have to be done at the time of N-release. However, this is 
between two bursts, where no giant \hbox{H\,{\sc ii}}-regions are present, 
providing no possibility for abundance observations by emission lines. 
Single bursts may produce scatter only if IMF 
parameters or enriched wind efficiencies vary from galaxy to galaxy. However, 
there is no indication of IMF variations, and the existence of enriched winds, 
on the whole, is questionable, and may be important in extreme low-mass 
galaxies only - see below. Double bursts produce the scatter quite naturally:
It was found reasonable above that a localized 
burst results in another burst, this time in the expanding supershell, 
surrounding the original burst. The timescale for star formation in the supershell
is found to be $\ga 2.4\times 10^7$ yr, comparable to the timescale for
O-ejection, but shorter than the timescale for N-ejection, hence producing the
desired scatter. 

In our model we assume that the very first burst is a single one, all other bursts 
appear in pairs. The
interburst period between the two bursts of a pair is tuned to give maximum
scatter, found below to be 30 Myr. The time between two pairs is set to 1 Gyr.

\subsection{The IMF} \label{imf}
The IMF used throughout this paper is the single-power Salpeter-IMF:

\BA
\phi (m)=\phi _0 m^{-2.35}
\EA
\noindent
\cite{salpeter:1955}.
$\phi _0$ is the normalization constant, found by $ \int _{m_L}^{m_U}m\phi 
(m)dm=1$. Using a lower mass cutoff $m_L=0.1 M_{\odot }$ and an upper mass 
limit $m_U=100 M_{\odot }$, $\phi _0$ is equal to 0.17. These limits are 
standard values, very often used in the literature. A Salpeter IMF still fits
the observations (above about $1 M_{\odot}$) quite well despite of its age 
and simple appearance \cite{leitherer:1998}.

The lower-mass cutoff is an important parameter because the yield from a
generation of stars is a function of the adopted lower mass. The reason is
that stars less massive than $\sim $1 $M_{\odot }$ do not eject metals, thus
locking-up all the material from which they are formed. Hence, lowering
the lower-mass cutoff implies a lower recycling fraction, giving a
lower yield.
As will be apparent from our results, it becomes attractive to invoke 
another value of $m_L$ namely 0.01 $M_{\odot }$. In this case the 
normalization constant is found to be 0.07.
Thus, the yield will be a factor 2-3 lower when using the lower 
cutoff.

Throughout this paper, a metallicity invariant Salpeter IMF has been used, 
since there are at present several observational indications of an abundance 
invariance \cite{wyse:1998,massey:1998}.
The number of stars with mass [$m_1$,$m_2$] is given by
\BA
N([m_1,m_2]) &=& \phi _0 M_{burst}\int _{m_1}^{m_2} m^{-2.35}dm \\ \nonumber
m_1  &=& m_j-\frac{\Delta m}{2}  \\  \nonumber
m_2  &=& m_j+\frac{\Delta m}{2}  \\ \nonumber
\EA
\noindent where $M_{burst}$ is the mass of a 
burst, i.e. the mass of gas turned into stars. The stellar mass grid may be 
as fine as one wishes. For use in the present 
model, a grid of 60 different stellar masses has been adopted, since this was found
to give sufficient mass resolution for the present purpose.

The mass of each burst may be constrained by the available observations of
SFRs. The SFR for BCGs is typically  0.1 - 1 $M_{\odot }\, 
\rmn{yr}^{-1}$.  For the mean SFR of a double burst to be within this range, 
the 
masses of the bursts may be estimated from $M_{burst}(1)+M_{burst}(2)/t_{ib}
\sim 0.1 - 1$, where $t_{ib}$ is the time between the two bursts of a pair. If
$t_{ib}=3\times 10^7$ yr, and assuming $M_{burst}(1)\sim M_{burst}(2)$,
this gives $M_{burst}\sim 1.5 - 15\, \times 10^6\, M_{\odot }$, in agreement 
with Marlowe et al. \shortcite[their table 7]{marlowe:1995}. 
\subsection{The equations}
The calculation of abundances is carried out just before every burst. The
reason for doing this is simply that we wish to compare with the observed
abundances, being measurable in \hbox{H\,{\sc ii}}-regions only. The adopted 
equations are described in the following.
The notation is similar to the one in Pilyugin \shortcite{pilyugin:1993}. 
The first step is to calculate the masses of each element present in the
gas phase just before every burst. For the $j^{th}$ burst:
\BA \label{M_i}
M_i(t_{j})&=&M_i(t_{j-1})-\Delta M_{j-1}X_i(t_{j-1}) \\ \nonumber
&-&\Delta W_i(t_{j-1})+\Delta W^{Inf}(\tau _{j,j-1})X_i^{Inf} \\
\nonumber
&+&\sum_{k=1}^{j-1} \Delta M_{k}(Q(\tau _{j,k})-Q(\tau _{j-1,k}))X_i(t_{k}) \\
\nonumber
&+&\sum_{k=1}^{j-1} \Delta M_{k}(Q_i(\tau _{j,k})-Q_i(\tau _{j-1,k}))
\EA
 
\noindent $t_{j}$ refers to the time just before the $j^{th}$ burst.
$X_i(t_{k})$ is the abundance of element $i$ just before the $k^{th}$ burst.
$\tau _{j,k}$ is defined as $t_{j}-t_{k}$, so it is the time elapsed since
the $k^{th}$ burst. Thus the element yields from stars with lifetimes
shorter than $\tau _{j,k}$ has to be included in the two summations.
$Q(\tau _{j,k})$ is the mass fraction of gas ejected from the $k^{th}$
generation of
stars, just before starburst $j$. The composition is left unchanged by stellar
nucleosynthesis, thus being the same as in the gas from which the stars
were formed. $Q_i(\tau _{j,k})$ is correspondingly the mass fraction ejected 
of element $i$, but newly synthesized. It is clear from this notation that 
$Q(\tau _{j,j})=Q_i(\tau _{j,j})=0$, since stellar lifetimes are finite. The 
first term on the right
side is the mass of element $i$ just before the $(j-1)^{th}$ burst. The second
term is the mass of element $i$ that has been turned into stars at burst $j-1$. 
The first summation term is the mass of element $i$ ejected, without being
changed, by the $k^{th}$ burst in the period between the $(j-1)^{th}$ and 
$j^{th}$ burst.
The second summation term is corresponding to the first one, except for the
fact that the mass of element $i$ was newly synthesized.
 
$\Delta W_i(t_{j-1})$ is the mass of element $i$, leaving the galaxy as a
result
of starburst $j-1$. This term consists of two parts: The part belonging to the
ordinary wind, and the one corresponding to the
enriched wind. Thus, the wind term may be written as
\BA \label{wind}
\Delta W_i(t_{j})&=&W_{ISM}\Delta M_{j}X_i(t_{j}) \\ \nonumber
&+&W_{SN}(Q_i(t_w)+Q(t_w)X_i(t_{j}))\Delta M_{j}
\EA
\cite{pilyugin:1993}.
$W_{ISM}$ and $W_{SN}$ are the efficiencies of the ordinary and enriched winds,
respectively. Both efficiencies are zero if the model is closed. The physics 
behind enriched winds is based on the principle that the supershell
following a burst, breaks up, allowing the hot ejecta to blow out and escape
from the galaxy. It is assumed that the wind is caused by supernovae type II, 
so the end of the wind phase $t_w$ is the lifetime of the least massive star 
exploding as a SNII, which is set to $8 M_{\odot }$, consistent with the 
mass adopted for our supershell calculations. Hence, for 
stars with masses above this limit, a fraction $W_{SN}$ of the ejected mass of 
element $i$ is leaving the galaxy. For stars less massive, $W_{SN}=0$. It is 
important to notice, in accordance with the physics
involved that this efficiency factor is the same for all elements considered,
but since oxygen is dominating the ejecta from SNII, the winds will be
enhanced in oxygen. 

The other possibility is ordinary galactic winds, arising as a consequence of
a general heating of the ISM, causing a fraction to leave the galaxy, i.e. the 
composition of the gas leaving the galaxy is the same as the composition of the
ISM. The first term of eq.\ \ref{wind} is the instantaneous-burst 
representation of Hartwick-outflow \cite{hartwick:1976}, giving a mass-loss 
proportional to the burst mass. By multiplying the mass of the wind with the 
fraction of the $i^{th}$ element in the ISM, taken just before the burst that 
is responsible for the wind, one obtains the mass removed of element $i$ due to 
an ordinary wind. $W_{ISM}$,$W_{SN}$ and the burst
masses are treated as free parameters, though one restriction is made, namely
that the yields of the first burst should be sufficient to place the second
burst {\it{approximately}} at the abundances of IZw18.

$\Delta W^{Inf}(\tau _{j,j-1})X_i^{Inf}$ is the increase in mass since the last
burst of element $i$ in the ISM due to inflow of gas. The composition of the 
infalling gas, given by $X_i^{Inf}$, is
taken to be primordial. The inflow rate is given by 
\BA \label{infmodel}
  \dot{M} (t)=\frac{M_0}{\tau _{inf}}e^{-t/\tau _{inf}}
\EA
\noindent \cite{lacey:1985,sommer:1993}. $M_0$ is the total mass accreted
for $t\gg \tau _{inf}$ and $\tau _{inf}$ is the accretion timescale. The
mass accreted at time $t$ is found by integrating eq.\ \ref{infmodel} from
0 to $t$ finding \mbox{$M(t)=M_0(1-e^{-t/\tau _{inf}})$}, finally giving
\BA \label{infcal}
  \Delta W^{Inf}(\tau _{j,j-1}) &\equiv &M(t_j)-M(t_{j-1}) \\ &=&M_0
  (e^{-t_{j-1}/\tau _{inf}}-e^{-t_{j}/\tau _{inf}})
\EA
\noindent
This is unfortunately
giving two free parameters further, namely $\tau _{inf}$ and $M_0$. In models 
not including inflow $\Delta W^{Inf}(\tau _{j,j-1})$=0.  
The mass of gas just before starburst number $j$ is
\BA \label{gas}
M_g(t_{j})&=&M_g(t_{j-1})-\Delta M_{j-1}-\Delta W(t_{j-1}) \\  \nonumber
&+& \Delta W^{Inf}(\tau _{j,j-1}) \\ \nonumber
&+& \sum_{k=1}^{j-1} \Delta M_{k}(Q(\tau _{j,k})-Q(\tau _{j-1,k})) \nonumber
\EA
\noindent  Finally the mass of the entire galaxy is
\BA \label{dwarf}
M_{dw}(t_{j})=M_{dw}(t_{j-1})+\Delta W^{Inf}(\tau _{j,j-1})-\Delta W(t_{j-1})
\EA
\noindent For the closed models $M_{dw}(t_{j})=M_{dw}(0)$.

$Q_i$ and $Q$ are adopted from models of stellar evolution as described
in section \ref{yield}. However, $Q_H(t)$ is calculated from
\BA
\sum_{i=H,He,C,N,O} Q_i(t)=0
\EA
\noindent assuming all newly consumed H to go into the production of
He, C, N and O. This simplification is realistic because the involved
elements are by far the most important.
 
From the above equations, it is now possible to calculate the abundances by
mass in the ISM, just before a new burst:
\BA
X_i(t_{j})=\frac{M_i(t_{j})}{M_g(t_{j})} \label{X_i}
\EA
\noindent
However, the observed abundances are not given by mass, but by number.  The 
abundance by number of, say $O(t_j)$ relative to $H(t_j)$ is
\BA
\frac{O}{H}(t_j)=\frac{M_O(t_j)}{M_H(t_j)}\; \frac{A_H}{A_O} \label{number}
\EA 
 
\noindent where A represent the atomic masses and M is the gas phase 
masses calculated by eq.\ \ref{M_i}.

In the calculations it is assumed that the ejecta mix into the
entire interstellar medium before the new burst appear. It should be noted
that this assumption is doubtful for the short interburst period between the
two bursts of a pair.
The fact that the metallicity of second generation stars is slightly
higher, is included in the sense that the adopted yields are metallicity
dependent.
The calculations are terminated if the amount of gas is insufficient
for a new burst, or the age of the dwarf galaxy is more than 15 Gyr.
 
The initial conditions to be put into the equations are the following.
The very first burst (the single burst) takes place at primordial gas 
composition, i.e. Z=0, $X_{He}$=0.243 \cite{izotov:1997}, $X_H$=0.757,
$X_C$=$X_N$=$X_O$=0. 

Before the first burst, the galaxy consists of gas only. Hence, the total mass
of the galaxy is equal to the mass of gas, in all calculations adopted to be
$10^8\; M_{\odot }$. However, the situation is a little different for the
inflow models. In this case, the initial mass of a galaxy is adopted to be 
zero. Gradually, the galaxy increases its mass by inflow of primordial gas, 
until the gas mass reaches a limit, sufficient to start the first burst. To be 
consistent, this mass limit has been adopted to be $10^8\; M_{\odot }$. So
the only difference from models not including inflow is that the 
first burst is delayed by the time it takes to build up a sufficiently large 
gas cloud.
\subsection{The adopted yields and their implementation} \label{yield}
Since the present models 
follow the evolution from the very first burst at Z=0 to about solar
metallicity, metallicity-dependent yields have been adopted. 
 The Renzini \& Voli \shortcite{RV:1981} yields for low- and intermediate mass
stars are the most widely used for the purpose of chemical evolution modelling.
Their strength is that they are giving the yields for several choices of
convection and wind parameters. Furthermore they give the primary yield
of nitrogen (and $^{13}C$) separately, which is very important for our 
purpose.
Indeed, one of the major problems to solve for low metallicity objects is the
extent of primary produced N, compared to the secondary production. Marigo,
Bressan \& Chiosi\ (1998) made also a separation between primary and 
secondary production in the mass interval 4-5$M_{\odot }$. The question whether
stars more massive than 5$M_{\odot }$ produce primary N is controversial.
The question is whether it is necessary at all to introduce such a source for
primary
N production. This is one of the questions we address in this paper.
The yields by van den Hoek \& Groenewegen \shortcite{van_den_hoek:1997}, also
for low- and intermediate mass stars, do {\bf not}
include separate results for primary nitrogen, so the yields given are a
mixture of primary and secondary components. This is the reason, why we have 
not adopted their yields for more than just a reference, see section \ref{result}. 
 
For use in the models, yields are selected in a consistent way, i.e. yields from 
Geneva tracks \cite{Maeder:1992} and yields from Padova tracks 
(Marigo, Bressan \& Chiosi\ 1996; Marigo et al.\ 1998; Portinari, Chiosi \& Bressan\ 1997)
 are kept in separate sets of 
yields. The adopted yields are organized as shown in table\ \ref{tabyield}.
\begin{table*}
\begin{minipage}{120mm}
\caption{The organization of the adopted yields.}
\label{tabyield}
\begin{tabular}{lcccc}
\hline
set no.  & Mass range ($M_{\odot }$) & Metallicities & Elements & Reference \\
1    & 9-100 & 0.001,0.02 & He,C,O,$m_{rem}$ & M92 (large mass-loss) \\
     & 9-40  & 0.002,0.02 & Nsec           & WW95\\
     & 0.8-8 & 0.001,0.004, & He,C,N, &    \\
     &       & 0.008,0.02 & O,$m_{rem}$ & HG97\\
     &       &                        &             &     \\
     &       &                        &             &     \\
2a   & 9-100 & 0.001,0.02 & He,C,O,$m_{rem}$ & M92 (large mass-loss) \\
     & 9-40  & 0.002,0.02 & Nsec           & WW95 \\
     & 1-8   & 0.004,0.02 & He,C,Nprim, &   \\
     &       &            & Nsec,O,$m_{rem}$ & RV81   \\
     &       &                        &             &    \\
     &       &                        &             &    \\
2b   &\multicolumn{4}{c}{As 2a, but small mass-loss for stars with
     $M>9M_{\odot }$} \\
     &       &                        &             &    \\
     &       &                        &             &    \\
3    & 6-100 & 0.0004,0.004, &  &   \\
     &       & 0.008,0.02 & He,C,Nsec,O,$m_{rem}$ &  P97 \\
     & 4-5   & 0.008,0.02 & He,Csec,Cprim, &  \\
     &       &     & Nsec,Nprim,Osec, & \\
     &       &       & Oprim,$m_{rem}$ & M98\\
     & 1.046-3 & 0.008,0.02 & He,C,Nsec,O,$m_{rem}$ & M96  \\
\hline
\end{tabular}

\medskip
Sources: M92:\cite[His tables 4,5 and 6]{Maeder:1992}, WW95:\cite[their tables
5A,5B,10A and 10B]{WW:1995}, HG97:\cite[their tables 13,17,35
and 38. Tables 13 and 17 are using a wind parameter $\eta $=4 for Z=0.008 and
Z=0.02. For Z=0.001 (table 38) and Z=0.004 (table 35), $\eta $ is 1 and 2
respectively]{van_den_hoek:1997}, RV81:\cite[their tables 3a,3d,3h and
3i. For all tables, $\eta $=0.333. For tables 3a (Z=0.02, Y=0.28) and 3h
(Z=0.004, Y=0.232), $\alpha $ is equal to 0. In the two other tables, $\alpha $
is 1.5, and $m_{HBB}$=8 $M_{\odot }$]{RV:1981}, P97:\cite[their tables 7 and
10]{portinari:1997},
M98:\cite[their tables 4 and 5, $m_{HBB}$=5 $M_{\odot }$]{marigo:1998} and
M96:\cite[their tables 4 and 5]{marigo:1996}
\end{minipage}
\end{table*}
\noindent 
Note that not all yields in sets 1, 2a and 2b are deduced directly from the Geneva 
tracks, as van den Hoek \& Groenewegen \shortcite{van_den_hoek:1997}, 
Renzini \& Voli \shortcite{RV:1981} and Woosley \& Weaver 
\shortcite{WW:1995} use different indirect methods.
 
The problem is to calculate the $Q$ and $Q_i$ terms just
before every burst, given the IMF and the ages of the previous bursts from 
which the stars eject the elements.
The mass ejected of an element $i$, at a time $\tau $ after a burst is
calculated as a sum of contributions from each star down to a stellar mass
$m_{\tau }$, corresponding to a lifetime $\tau $. The lifetime of each star is 
found by linear Lagrange interpolations in both mass and metallicity between 
the values given by the stellar tracks. 
The following expression is used for calculating $Q_i$
\BA
Q_i=\sum_{m_j>m_{\tau }}q_i(m_j)\: N(m_j)
\EA
\noindent where $N(m_i)$ is the number of stars with mass $m_i$, found with the
adopted IMF, and $q_i(m_i)$ is the mass ejected of element $i$ from stars 
having this mass. The $q_i$s are found from the references in table\ 
\ref{tabyield} in two steps.
First, a linear interpolation is performed between the values for the available
metallicities. If the metallicity is lower than the metallicities, for which 
the yields are known, primary and secondary components of N are treated 
differently.
The yields for primary elements are taken to be equal
to the yields at the lowest metallicity with known yields, except for set no. 
3, in which the primary N is obtained by extrapolating
linearly from the two metallicities with known yields. For this set, an
extrapolation should be more correct than just assuming the yield to be
constant below the range of metallicities covered by the yield sources,
since the range does not extend below Z=0.008 for the critical mass interval
4 - 5 $M_{\odot }$. The price to pay is an extrapolation reaching far
beyond the covered metallicity range. Note that stars with masses less than 4 
$M_{\odot }$ or larger than the upper mass for hot bottom burning (free 
parameter in
the model for set 2a and 2b, and 5 $M_{\odot }$ for set 3) are only producing
secondary N. The secondary N yield is interpolated between the 
lowest metallicity with known yields and Z=0, since the yield of secondary N 
is 0 at Z=0, according to the definition of secondary production.
For H, He, C and O  the yields interpolated/extrapolated are the sum of primary +
secondary yields. This is without influence on the results.

Second, the yields are interpolated linearly with respect to initial
stellar mass, finally giving the $q_i$s. 
For 2a and 2b, the mixing length parameter $\alpha $, defined as the convection
mixing length divided by the pressure scale height, is a free parameter. We
know the yields for two different values of $\alpha $ from Renzini \& Voli
\shortcite{RV:1981}, namely $\alpha $=0 and $\alpha $=1.5. Thus
when implementing the yields, the $q_i$s are found in three steps. Before
the two steps described above are carried out, the yields given in the
mass interval 4 $M_{\odot }$ to the upper mass of HBB, $m_{HBB}, 
$variable between 5 and 8 $M_{\odot }$, are linearly interpolated with
respect to $\alpha $.
 
The implementation of the mass ejected unprocessed is very similar.
The expression for the total mass ejected since the starburst is written as
\BA \label{return}
Q=\sum_{m_i>m_{\tau }}(m_i-m_{rem}) N(m_i)
\EA
\noindent $m_{rem}$ is the mass of the stellar remnant (white dwarf, neutron
star or black hole). Again the value of $m_{rem}$ is extracted from the
sources by linear interpolation, first with respect to metallicity, then
between initial stellar masses.
 
It is very useful to calculate the true yields of selected elements because
they allow for an easy comparison between the nucleosynthetic outcome of
different stellar models.
The true yield of an element is defined to be the mass of an element
ejected from a star generation, divided by the fraction locked
up in stellar remnants and low-mass stars.
\begin{table} 
\caption{The true yields and return fractions for the sets in use.}
\label{totalyield}
\begin{tabular}{lcccc}
  & Z & $p_{He}$ & $p_{O}$ & R \\
Set 2a & 0.001 & 0.032 (0.011)& 0.020 (0.007) & 0.274 (0.116)\\
       & 0.02 & 0.031 (0.011)& 0.008 (0.003) & 0.283 (0.120)\\
       &      &               &               &  \\
Set 2b & 0.001 & 0.032 (0.011)& 0.020 (0.007) & 0.274 (0.116)\\
       & 0.02 & 0.024 (0.008)& 0.014 (0.005) & 0.282 (0.119)\\
       &      &               &               &  \\
Set 3  & 0.001 & 0.076 (0.026)& 0.014 (0.005) & 0.279 (0.118)\\
       & 0.02  & 0.096 (0.032)& 0.014 (0.005) & 0.299 (0.127)
\end{tabular}

\medskip

The values given are for the high low-mass cutoff. Numbers in brackets are for 
the low cutoff. Note that the He yields for set 3 are higher than for the 
other sets by a factor of about 3.  The O yields are generally quite robust, 
though the yields of set 2a at solar metallicity (Z=0.02) are lower by a factor 
of about 2, compared to the other O yields at this metallicity. R is the return
fraction.
\end{table}
\noindent
The total mass ejected of an element, e.g. O is given by $\sum_{m_i}q_O(m_i)\: 
N(m_i)$, where $q_O(m_i)$ is found from the sources following the proces
described above. The lock-up fraction is equal to
$1-$R, where R is the return fraction, i.e. the mass ejected in units of the
total mass of the burst $Q/M_{burst}$.
The resulting true yields are given in table\ \ref{totalyield} for
He and O, as we shall need these quantities later. Note that the true
He yields for set 3 are higher than those of 2a and 2b, in particular at
the higher metallicities. Note also the lower yields, when using
$m_L$=0.01, by a factor of $\sim $2.8.

\subsection{Stellar lifetimes and element timescales} \label{timeofeject}
In the case of set no. 1, 2a and 2b, the stellar lifetimes are adopted
from Schaller et al. \shortcite{Schaller:1992} (Geneva-tracks), while in
set no. 3 they are from Portinari et al. \shortcite{portinari:1997}
(Padova-tracks). To compare the two sources, lifetimes are shown for
some selected metallicities and masses in table\ \ref{lifetime}.
\begin{table}
\begin{centering}
\caption{The stellar lifetimes for the two stellar tracks.}
\label{lifetime}
\begin{tabular}{rcccc}
\hline
 M & \multicolumn{2}{c}{Geneva} & \multicolumn{2}{c}{Padova} \\
$M_{\odot}$ & Z=0.001 & Z=0.02 & Z=0.0004 & Z=0.02 \\
100 & 3.21E+06 & 3.09E+06 & 3.38E+06 & 3.39E+06 \\
60  & 4.08E+06 & 3.89E+06 & 4.19E+06 & 4.12E+06 \\
40  & 5.34E+06 & 4.79E+06 & 5.44E+06 & 5.12E+06 \\
20  & 1.02E+07 & 8.96E+06 & 1.05E+07 & 9.15E+06 \\
15  & 1.45E+07 & 1.28E+07 & 1.52E+07 & 1.33E+07 \\
6   & 1.00E+08 & 1.08E+08 & 7.62E+07 & 7.45E+07 \\
4   & 3.41E+08 & 4.40E+08 & 1.82E+08 & 2.03E+08 \\
2   & 1.99E+09 & 2.91E+09 & 1.08E+09 & 1.21E+09 \\
1   & 1.07E+10 & 1.74E+10 & 7.13E+09 & 1.03E+10 \\
\hline
\end{tabular}
\end{centering}
\end{table}
\noindent It is possible directly to compare the stellar lifetimes at solar 
metallicity. Low- and intermediate mass stars seems to have shorter lifetimes 
using the Padova-tracks (between $\sim $40 to $\sim $70 per cent), while the 
opposite is true for massive stars (but $\la $10 per cent longer). This age
difference may be a result of different stellar wind-efficiencies, and
the inclusion of overshooting in the Padova-tracks.

By combining our knowledge of stellar yields with the respective lifetimes as 
function of mass, we obtain information on the ejection timescales of 
different elements. For the present purpose, it is the difference between the 
timescales of N and O ejection that is particularly interesting, since the time when 
the {\underline{ejected}} N/O has its minimum is equal to the time, where the 
second burst of a pair should appear if maximum scatter is to be obtained. The 
timescales are found for a fixed metallicity at which the ejected masses of O 
and N of one burst are calculated in small timesteps (of the order of $\sim 
10^6$ yr) until all stars down to 1 $M_{\odot }$ have ended their life cycle.
The time evolution of the element ejection has been calculated for set no. 2a and 
3 for two metallicities. 
Fig.\ \ref{timeRV} shows the timescales for set 2a. The results for set 3 are 
very similar.
\begin{figure}
\epsfxsize=\linewidth
\epsfbox{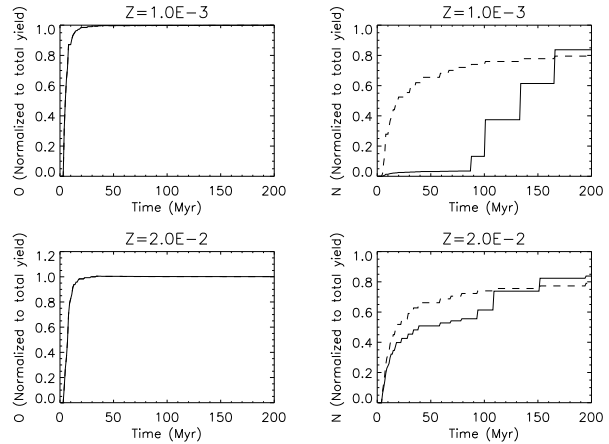}
\caption{The time evolution of O and N ejection after an instantaneous burst 
using set 2a, at two different
metallicities, and two different values of $\alpha $. Dashed lines correspond
to $\alpha $=0, solid lines to $\alpha $=1. O is to the left and N to the
right. The y-axis is the mass ejected of N or O,  in units of the total mass 
ejected of N or O when all stars have ended their life. The step-like appearance 
of the curves is a consequence of the combined action of discrete stellar masses 
and finite timesteps. $t$=0 is the time of the burst.} \label{timeRV}
\end{figure}
 
The production of N in massive stars is different for the two metallicities. 
For solar metallicity, it is released much faster for $\alpha$=1, i.e by more 
massive stars. 
This is only secondary N though, since the contributors have lifetimes shorter 
than 50 - 80 Myr, hence being more massive than 8 $M_{\odot }$ (compare with 
the stellar lifetimes given in table\ \ref{lifetime}). For set 2a, it is not 
surprising that the ejection timescale of N is much shorter at low metallicity
if $\alpha $=0.0 (dashed lines in fig.\ \ref{timeRV}) than if $\alpha $=1.0, 
because no primary nitrogen is ejected from intermediate mass stars when 
$\alpha $=0.0. The major N production sets in at about 100 Myr at the low 
metallicity. For solar metallicity a large fraction is produced within the 
first 50 Myr, about 40 (2a) to 60 per cent(3). Almost all of the oxygen is 
produced in $\la $30-40 Myr for all sets and metallicities.
 
Hence, in order to choose the interburst period between the bursts
of a pair, a minimum value of N/O would appear somewhat between 30 and
80 Myr for the low metallicity and between 30-40 Myr for the high metallicity.
If a choice has to be made, 30 Myr would be the best to make, since, in
general, it is easier to reproduce scatter at low metallicities due to
the relative effect of ejected metals to metals already present in
the interstellar medium. Following this conclusion, the time between the
two bursts of a pair is set to 30 Myr.

\section{Results} \label{result}
The model is applied to fit the observations in the N/O-O/H plane with the use
of the yields in table\ \ref{tabyield}. The accompanying results on Y-O/H and 
O/H vs. gas fraction are displayed as well. 
The order of presentation starts with the results of the closed model.

\subsection{The closed model}
The closed model has been applied to all of the yield sets.
In most plots, the IMF with $m_L$=0.1 has been used. For those employing
the 0.01 low-mass limit, it is mentioned separately in the text. It is not
necessary to do all calculations for both IMF's, since the results are the
same, except for the gas fractions. For all elements, the only difference is a
lower yield using the low value of $m_L$,
hence more massive bursts are needed to produce the same abundances as when
the higher low-mass limit is used. This statement is confirmed in the
following.

\subsubsection{N/O-O/H}
\begin{figure} 
\centering
\epsfxsize=6cm
\epsfysize=4cm
\epsfbox{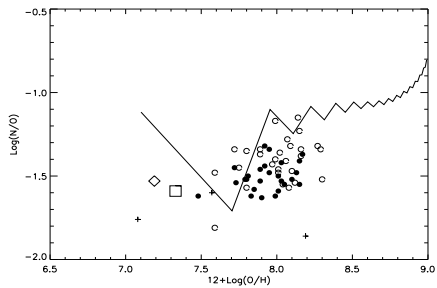}
\caption[]{Results for a closed model using yield set no. 1 and 
$m_L=0.1\: M_{\odot }$ The following
parameters were adopted: Period between pair of bursts is 1 Gyr and the short
interburst periods are 30 Myr. The mass of the first burst is
$1\times 10^6$ $M_{\odot }$ and the masses of the successive bursts are
$3\times 10^6$ $M_{\odot }$. \\
The observations shown are the same as in fig.\ \ref{sampleNO}.} \label{HG1}
\end{figure}
 
Fig.\ \ref{HG1} shows the evolutionary path of a double-bursting 
dwarf galaxy, as it has been calculated by the model using yield set 1. The 
point where the evolutionary path starts
(at 12+log(O/H)$\sim $7.1) is where the second burst appears, 1 Gyr after the
first burst. The next point is where the third burst would be observed, 30 Myr
after the second burst. Thus, the upper points of the saw-tooth pattern are
corresponding to the first bursts of each pair, and the lower points to the 
second pair-bursts. The scatter is
more pronounced at low metallicities because the mass of O ejected relative
to the one already present in the ISM is higher, the lower the metallicity.
It is clear from this figure, why set no. 1 is used as a reference only. The 
level of N/O is much too high, in particular at low metallicities. The 
only way one can lower the ratio is by assuming a top heavy IMF, since the
yields from van den Hoek \& Groenewegen \shortcite{van_den_hoek:1997} do not
allow for changing the convection parameters or provide possibilities for
distinguishing between primary and secondary components. Thus, it is not
possible to scale the secondary components separately to Z=0. Further, neither
outflow nor inflow is able to cure the problem, in particular not if
selective winds are considered, since this would increase N/O even more.
 
It is interesting to note that the parameters (convection, wind etc.) used
to fix the yields in van den Hoek \& Groenewegen \shortcite{van_den_hoek:1997}
have been found by using a synthetic model on AGB stars in the LMC and in the
Galactic disc. The LMC is not represented in fig.\ \ref{HG1}, but according
to Pagel et al. \shortcite{pagel:1992} 12+log(O/H)= 8.36 and log(N/O)= -1.22 
for LMC, which is fitted fairly well after 4 pairs of bursts (but may be fitted
by 2 pairs of bursts if the mass of each burst is larger), thus being
consistent. The problem seems to arise at lower metallicities.

Next, set 2a is applied (fig.\ \ref{RV1}). The upper mass limit of hot bottom 
burning (HBB) is set to $m_{HBB}=5$$ M_{\odot }$, compatible with set no. 3. In this 
case, a value of $\alpha $=1.1 is necessary to obtain the right level of N/O. 
A larger value of the convection parameter would produce too much primary N 
and vice versa.
\begin{figure}
\epsfxsize=6cm
\epsfysize=4cm
\epsfbox{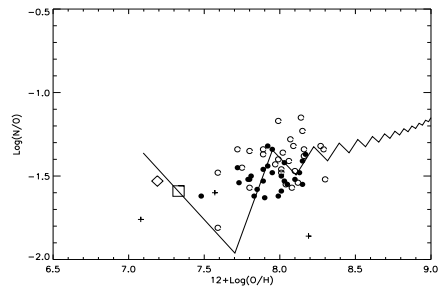} \vfill
\caption[]{Results for a closed model using yield set no. 2a and
$m_L$=0.1. The following parameters are used: Period between pair of
bursts is 1 Gyr and the short
interburst periods are 30 Myr. The mass of the first burst is
$1\times 10^6\: M_{\odot }$ and the masses of the successive bursts are
$3\times 10^6\: M_{\odot }$. Finally, the upper mass of hot bottom burning,
$m_{HBB}$, is $5 M_{\odot }$. For this plot,
$\alpha  =1.1$ is used, since this value places N/O at the right level.} \label{RV1}
\end{figure}
\begin{figure}
\epsfxsize=6cm
\epsfysize=4cm
\epsfbox{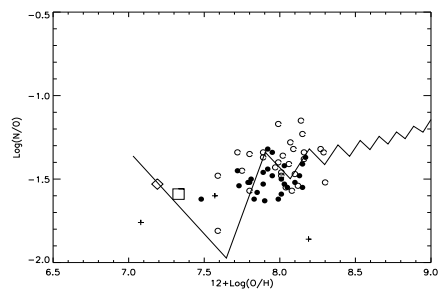}
\caption{Same yield set and parameters as in fig\ \ref{RV1},
except that $m_L$=0.01, the mass of the first burst is $2\times 10^6\: M_{\odot
}$ and the masses of the following bursts are $6\times 10^6\: M_{\odot }$. 
Compare the results to those of fig\ \ref{RV1}. It is evident that the observations are
explained equally well using the low cutoff as when using the higher cutoff, if just 
burst masses are about twice as massive.}
\label{RVNOlow}
\end{figure}
Both theoretical predictions and observations of AGB stars in the Large 
Magellanic Cloud are suggesting a level of HBB, roughly corresponding to 
$\alpha $=2.0 \cite[and references therein]{van_den_hoek:1997}, which is well 
above the value obtained here. If $\alpha $ really is so large in AGB stars in 
the LMC, an explanation might be that the value of the convection parameter 
depends on metallicity, since the metallicity of the LMC is high (about 
half-solar). A higher value of $\alpha $ at high metallicities is not ruled out
by the model as long as it is equal to 1.1 during the first few
bursts.

A large spread is evident and sufficient to explain the observed scatter.
Here, the plusses, representing the DLA-systems, are neglected. A similar run is
presented in fig.\ \ref{RVNOlow}, only employing the IMF having $m_L$=0.01 
$M_{\odot }$. As expected, it is seen that the result resembles
the situation in fig.\ \ref{RV1}, if one assumes all bursts to involve twice
the mass of the bursts in fig.\ \ref{RV1}.

Figures\ \ref{RV1} and \ref{RVNOlow} give a good comprehension of the outcome
of the model, but some considerations should be made before any solid 
conclusions are drawn. Firstly, abundances are measured in \hbox{H\,{\sc 
ii}}-regions only, and therefore only at the
'positions' of the bursts, not in between. Secondly, the evolutionary path
represents the evolution of one dwarf galaxy. Another dwarf galaxy, having
different parameters such as total mass, burst masses etc., would have
another evolution, and thus another path in the N/O-O/H plot. Thus, fig.\
\ref{RV2} shows the evolutionary paths of dwarf galaxies having different
fractional burst masses. Yield set 2a has been used, and
all of the parameters, except the mass of each burst, are the same.
\begin{figure}
\centering
\epsfxsize=\linewidth
\epsfbox{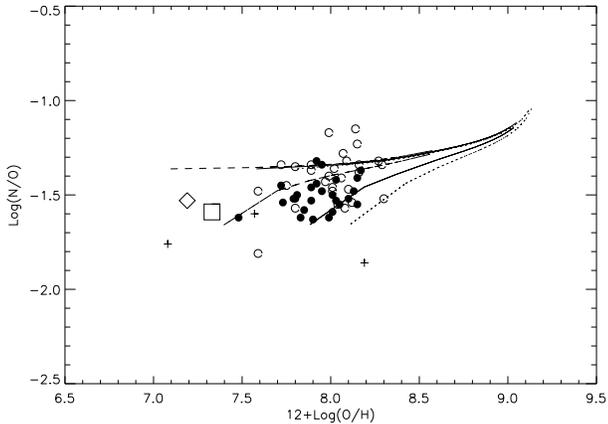}
\caption[]{Three closed models, each with different burst masses, using set 
no. 2a. All other parameters are the same as in fig.\ \ref{RV1}. For
each burst mass, two lines have been drawn: one through the upper points of
the saw-tooth pattern and one through the lower points. The dashed lines
correspond to burst masses equal to 1 per cent of the total mass of the galaxy, 
the solid lines to 3 per cent and the dotted lines to 5 per cent} 
\label{RV2}
\end{figure}
To make the plot more clear, lines have been drawn through the first bursts of 
each pair, and through the second bursts. The former lines are almost coincident, 
because the time interval between two pairs 
of bursts is $\sim $ 1 Gyr, enough to release almost all of the nitrogen. Thus, 
larger bursts increase both N and O, adjusting N/O to be roughly unchanged. 
The case is different for the lines through the second pair-bursts, because 
the N from the first pair-burst has not yet been released after 30 Myr. Thus, 
a larger burst mass gives an increase in O abundance only.
 
Two results may be deduced from fig.\ \ref{RV2}. Firstly, the scatter of the
observations is perfectly explained, because an observational scatter of
$\sim $0.1 dex has to be taken into account. Furthermore, the area between
the lower line corresponding to 5 per cent burst masses and the upper line is 
'filled' in the sense that the distribution of burst masses may be everything 
between zero and 5 per cent to match the observations. Thus, the distribution of 
abundances is accounted for by using the models of dwarf galaxies 
having various reasonable relative burst masses. The other result is the slight
increase of the evolutionary path toward higher metallicities. However, the 
small upturn may not be a consequence of dominating 
secondary N production, but rather the decreasing O yield at higher
metallicities, see table\ \ref{totalyield}. 
\begin{figure} 
\centering
\epsfxsize=\linewidth
\epsfbox{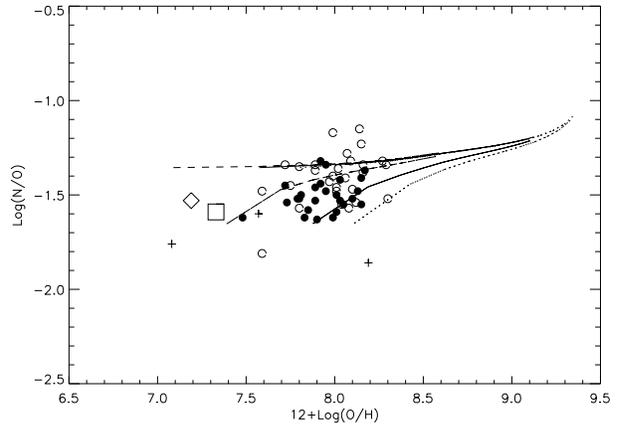}
\caption[]{Evolutionary paths of three closed models using yield set 2b. The lines are
drawn as in fig.\ \ref{RV2} with identical parameters so the dashed lines
correspond to burst masses equal to 1 per cent of the total mass of the galaxy, 
the solid lines to 3 per cent and the dotted
lines to 5 per cent} \label{RV2b}
\end{figure}
 
Set no. 2b is different from no. 2a only in that a modest mass-loss from
massive stars is assumed. The effect of this change on the evolutionary
path is seen by comparing fig.\ \ref{RV2b} with fig.\
\ref{RV2}. It is possible directly to compare the model outputs of the
two figures, since all parameters, including the burst masses, are
identical. When using set 2b instead of 2a, the N/O ratio has an even smaller
upturn at higher metallicities, because the O-yield is higher. Thus, even 
when N production is increasing at higher metallicities, so is the oxygen, 
keeping the ratio down and thereby hiding the secondary behavior of N. Still, 
the conclusions are the same: both the level of N/O and the scatter
is explained, independent of $m_L$.

Using yield set no. 3, the same kind of plot is constructed as those
above. Here there is no explicit assumption on the convection parameter, but
for reasons explained below, the primary component of
N (and O) for Z$< $0.008 has been extrapolated linearly, using the yields of
Z=0.008 and Z=0.02 instead of just adopting the primary yield at Z=0.008.
The run of the model is presented in fig.\ \ref{pad}.
\begin{figure} 
\centering
\epsfxsize=\linewidth
\epsfbox{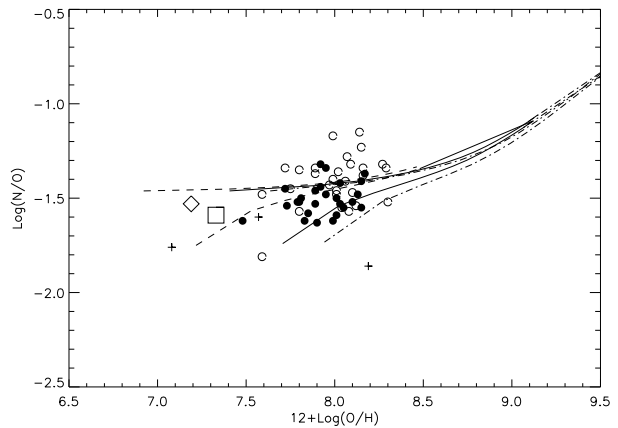}
\caption[]{Three closed models each with different relative burst masses using
the yields of set 3 (Padova). The meaning of the lines is the same as in
fig.\ \ref{RV2}. The dashed line corresponds to burst masses equal to 1 per 
cent of the total mass of the dwarf galaxy, the solid line to 3 per cent and 
the dashed-dotted line to 5 per cent. It is worth
noting that the mass interval for producing primary N is the same as in
fig.\ \ref{RV2}. For these runs, all primary yields have been extrapolated for
metallicities lower than the range covered by the yield sources.} \label{pad}
\end{figure}
At least two features have to be mentioned. The first is that the scatter is 
somewhat smaller than for set 2a (or 2b) and the second is an upturn in N/O at
higher metallicities, being more pronounced, than it was for set 2a and 2b.
Unfortunately, the upturn is taking place for higher oxygen abundances than 
those of the observations. 

The primary component in set 3 has also been modelled in another way: For 
the same choice of parameters as in fig.\ \ref{pad} (solid line), the model has 
been applied with primary yields equal to the one at the lowest metallicity 
for which the yields are given (Z=0.008), with the resulting evolutionary path
plotted in fig.\ \ref{pad1}. A comparison shows that an adoption of
the primary yields at Z=0.008 will produce too little primary N. It seems 
to be necessary to make primary production more effective, the lower the
metallicity, as it is done by the linear extrapolation, because the primary
N yield at Z=0.008 is slightly larger than it is for solar metallicity. This
is actually an important result, because it provides a constraint
on the produced primary N, independent of stellar models. The explanation
for this result might be that the number of thermal pulses in the AGB phase of
low metallicity stars is higher, thus converting more C and O into N via the
CNO-cycle. Hence, in the following, all model runs using set 3, will use 
linear extrapolations of primary N yields, below Z=0.008.
Of course, one could make the IMF steeper thereby producing less oxygen.
However, observations do not support a steeper IMF.

When comparing the outputs of the model, using set 2a (2b) and 3, it
is important to remember that in set 2a and 2b, the lowest metallicity for which
yields are available is Z=0.004, at least in the critical, and very important
mass interval 4 to 5 $M_{\odot }$, while it is as high as Z=0.008 in the
same mass interval for set 3. Thus, whether primary N yields at Z $<$0.008 are 
extrapolated from Z=0.02 over Z=0.008 or are adopted to be equal
to the yield at Z=0.008, the result is quite uncertain, and stellar yields
at much lower metallicities are definitely desirable.
\begin{figure}
\centering
\epsfxsize=\linewidth
\epsfbox{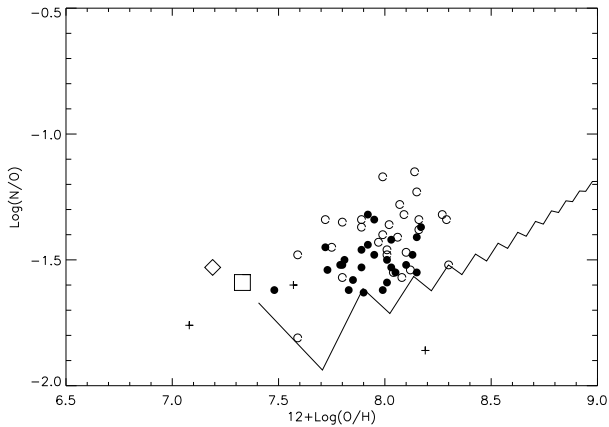}
\caption[]{An evolutionary track, using set 3 yields. All burst masses are 
$3\times 10^6\: M_{\odot }$. The yields of primary components for
metallicities lower than Z=0.008 are set equal to those given for Z=0.008.}
\label{pad1}
\end{figure}

\subsubsection{Y-O/H} \label{Y}
It is worth emphasizing that the helium abundances are fitted with input 
parameters not different from those used to fit the N/O ratio.
\begin{figure}
\centering
\epsfxsize=\linewidth
\epsfbox{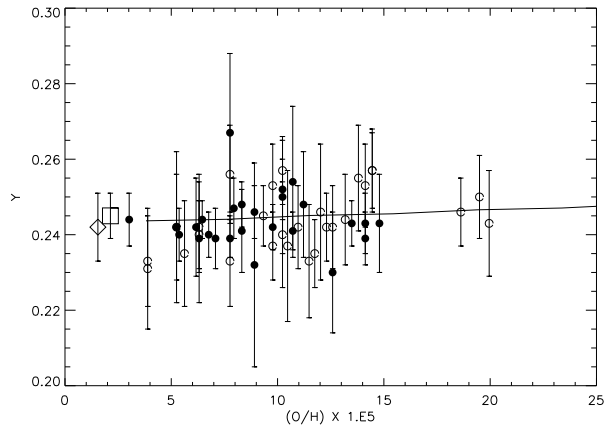}
\caption{The evolution of the helium mass fraction Y as a function of O/H, 
using yield set no. 2a and $m_L$=0.1, plotted as a solid line. The 
model output of Y and O/H are from the same run as marked
by the solid line in fig.\ \ref{RV2}, thus having a burst mass of 3 per cent of 
the total galaxy mass. The meaning and sources of the symbols are the same as 
in the N/O - O/H plots, except that the DLA systems are not represented. Note
that the O/H axis is not logarithmic.} \label{RVY}
\end{figure}
For the closed model presented in fig.\ \ref{RV2} by the solid line, Y has been
calculated as well. Y against O/H is shown in fig.\ \ref{RVY}.
The evolutionary Y path does not show any scatter, though the dwarf galaxy is 
double bursting, because about half of the helium is produced in massive stars, 
together with the oxygen. 
The fit is very good and well within the error bars. For the two most
metal deficient objects in the plot, IZw18 and SBS0335-052, it seems
to be necessary to assume smaller burst masses, say 1 per cent of the total 
mass - the model with the dashed lines in fig.\ \ref{RV2}. Thus, these
objects can be explained by experiencing their second burst and having
burst masses equal to 1 per cent of their total mass. The output of the model 
gives the abundances just before every burst. Hence, to calculate the 
model value of dY/dZ, the Y,O/H values were fitted using a linear least-square
fit, assuming Z=20(O/H). The result is 1.0,
lower than the (very uncertain) value of 2.6 derived from the observations, see
section\ \ref{abundance}. The corresponding result using set 2b is not shown, 
since it resembles the one by set 2a. Certainly, it is true that the He yield 
is smaller for modest mass loss, but only for very massive stars and high
metallicities \cite{Maeder:1992}.

The dependency of He on O is more pronounced, when using
the Padova yields, as seen in fig.\ \ref{ypad}.
\begin{figure}
\centering
\epsfxsize=\linewidth
\epsfbox{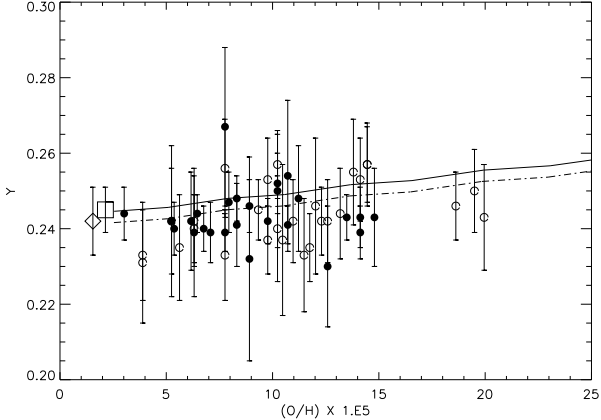}
\caption[]{The evolution of Y using the Padova yields.
The solid line is the evolutionary path, using a primordial He abundance
$Y_p$=0.243, from the same model run as used for the solid line in
fig.\ \ref{pad}, and the
dashed-dotted line is the corresponding path, assuming $Y_p$=0.240, still
within the uncertainties of the primordial He determination. The mass of
each burst is 3 per cent of the total mass in both cases.} \label{ypad}
\end{figure}
Calculating dY/dZ, using this yield set, in the same way as done for set 2a
one obtains 3.1, higher than the value calculated for the observations (2.6).
From table\ \ref{totalyield} it is clear that the true He yields for set 3
are a factor of $\sim $3 higher than those of sets 2a and 2b, but it is
unclear why the Padova tracks produce that much helium.
This feature makes the fit not quite as good as it was for set 2a (2b).
However, the primordial helium determination is only certain within 0.003
\cite{izotov:1997}. Setting the primordial He abundance equal to the lower
limit (0.240) in the initial conditions of the model does not change the
N/O-O/H evolutionary path, but it does lower the level of Y in fig.\
\ref{ypad}. The new evolutionary path of Y has the same slope as the solid one 
of course, but sufficiently lower to be a fairly good fit, as it appears well 
within the observational uncertainties.

\subsubsection{O/H - $\mu $} \label{gasmetal}
Gas fractions may be useful in serving as a further constraint on chemical 
evolution models. It has
been pointed out by several authors that closed models, such as the one 
presented above, are unable to explain the observed gas fractions 
\cite{mc:1983,mt:1985,carigi:1995}. Thus, gas fractions have been calculated 
just before every burst using eqs.\ \ref{gas} and \ref{dwarf} allowing for
comparison with some observed values from the literature.

Problem lies in the interpretation of the observations. The total dynamical 
mass of a dwarf galaxy also includes dark matter (DM). This dark is unlike to 
participate in the chemical evolution.
Our model is including baryonic matter only. Thus, in
order to compare the observations with the calculated gas fractions, one has to
be sure that no significant amount of non-baryonic DM is
included in the total mass estimates. However, as pointed out by several
authors \cite{brinks:1988,kumai:1992,carigi:1998,meurer:1998} dwarf 
galaxies, including BCGs, are DM-dominated. Consequently, dynamical 
estimates of total masses are useless, when considering dwarf galaxies. 
Carigi et al. \shortcite{carigi:1998}
tried to solve the problem by using dynamical estimates within the visible  
Holmberg radius, to avoid inclusion of a DM halo. Unfortunately, this leaves us
with a very small sample of observations, and still the amount of DM within the
Holmberg radius is uncertain.
 
However, gas fractions are one of the few possible ways of constraining chemical
evolution models, so a discussion of gas fractions, using the same models as in 
the preceding paragraphs should be included. To avoid erroneous conclusions, we
used lower and upper limits of the gas fractions.
 
The gas fractions used are presented in table\ \ref{gasfrac}. The gas masses
included are HI-masses multiplied by 1.3 to account for the helium content.
The amount of $H_2$ is ignored. All dynamically estimated gas fractions have 
been scaled to a Hubble constant $H_0=65\, \rmn{km}\, \rmn{s}^{-1} 
\rmn{Mpc}^{-1}$ using 
$M_{gas}\propto (distance)^2$ and $M_{tot}\propto (distance)$, so the gas 
fraction $\mu \propto (distance)\propto H_0^{-1}$.
 
DM is implicitly included, and molecular hydrogen is ignored, hence these
gas fractions must be regarded as lower limits. Unfortunately, even the lower
limits seem to be very uncertain. For instance, the gas fraction of IZw18 was 
first calculated using the data from Staveley-Smith, Davies \& Kinman\ (1992). 
However, after He-correction and $H_0$-scaling, the
gas fraction became larger than 1. This is also the case for some of the
galaxies in the Thuan \& Martin \shortcite{Thuan:1981} sample (not included in 
the
present sample). In the case of IZw18, the value 15 per cent gas has been 
adopted from Matteucci \& Chiosi \shortcite{mc:1983}. 
\begin{table*}
\begin{minipage}{120mm}
\caption{The data used for gas fraction plots.}\label{gasfrac} 
\begin{tabular}{lccccc}
\hline
 Name & $L_B\; (L_{\odot })$ & $M_{gas}\; (10^8\,M_{\odot })$  &
$log\left( \frac{M_{gas}}{M_{tot}}\right)$ & $log\mu $ &
$12+log(O/H)$\\
 & (1) & (2) & (3) & (4) & (5) \\
Mrk 5 & 0.35E+9 & 2.25 & -2.65E-3 & -0.13 & 8.21\\
\hspace{20pt}36 & 0.96E+8 & 0.33 & -5.03E-3 & -1.19 & 7.85\\
\hspace{20pt}67 & 0.17E+9 & 0.28 & -1.07E-2 & -0.52 & 8.21\\
\hspace{20pt}178 & 0.96E+8 & 0.24 & -6.82E-3 & -0.09 & 7.95\\
\hspace{20pt}71 & 0.11E+10 & 17.1 & -1.16E-3 & -0.01 & 7.89\\
\hspace{20pt}209 & 0.76E+8 & 0.87 & -1.52E-3 & -0.38 & 7.81\\
\hspace{20pt}600 & 0.44E+9 & 5.71 & -1.34E-3 & -0.68 & 8.01\\
IZw18  & 0.10E+9 & 1.12 & -1.63E-3 & -0.83 & 7.22\\
IIZw40 & 0.23E+9 & 7.79 & -5.04E-4 & -0.29 & 8.14\\
IIZw70 & 0.15E+10 & 5.88 & -4.31E-3 & -0.54 & 8.06\\
VIIZw403 & 0.53E+8 & 0.66 & -1.40E-3 & -0.39 & 7.73\\
IZw49 & 0.17E+10 & 7.96 & -3.76E-3 & -0.28 & 8.03\\
IZw123 & 0.36E+9 & 0.85 & -7.31E-3 & ? & 8.10\\
CG 1116+51 & 0.15E+9 & 3.63 & -7.00E-4 & -0.07 & 7.52\\
SBS 335-52 & 0.98E+9$^e$ & 1.94$^e$ & -8.80E-3 &  ? & 7.33\\
NGC 6822 & 0.18E+9$^a$ & 2.53$^a$ & -0.13$^d$ & -0.80 & 8.19\\
SMC & 0.99E+9$^b$ & 4.66$^c$ & -0.31$^d$ & -0.38$^a$ & 8.04\\
LMC & 0.47E+10$^b$ & 6.90$^c$ & -0.65$^d$ & -0.92$^a$ & 8.36\\ \hline
\end{tabular}

\medskip
(1) The blue
luminosities are from Thuan \& Martin \shortcite{Thuan:1981} with four exceptions,
see (a,b,e). The luminosities are scaled to h=0.65.
(2) The gas masses are from Thuan \& Martin, though see (a,c,e). All values are
scaled to h=0.65.
(3) The logarithm of the calculated upper limit gas fractions, where
$M_{tot}=M_{gas}+M_{stars}$. $M_{stars}$ has been calculated assuming the
galaxies to do their very first burst, see text and note (d).
(4) The logarithm of the observed, lower limit gas fraction. $\mu $ is the
gas mass divided by the dynamically estimated total mass from Thuan \&
Martin (1981), Matteucci \& Chiosi \shortcite{mc:1983} and Taylor et al.
\shortcite{taylor:1994}. The question marks indicate that the gas fraction is 
not found, so in these cases, only the upper limits are used. All values are
scaled to h=0.65, though see note (a).
(5) The oxygen abundances from Kobulnicky \&
Skillman \shortcite{kobul:1996}, Pagel et al. \shortcite{pagel:1992} and Izotov 
et al. \shortcite{izotov:1997}.  \\
(a) Values are from Lequeux et al. \shortcite{leq:1979}, obtained by 
h-independent
distance measurements. Hence, these values are adopted without any scaling. \\
(b) These luminosities are calculated from the absolute magnitudes given in
Matteucci \& Tosi (1985), using the solar absolute blue magnitude 5.48. \\
(c) Values from Carigi et al. \shortcite{carigi:1998}. They give the gas masses
directly for h=0.65 and account for $H_2$. To be consistent with the rest of
our sample, their values are divided with 1.1 to obtain gas masses not 
including molecular hydrogen.\\
(d) These galaxies are known for sure not to experience maximum luminosity of
their first starburst at the present epoch. Thus,
more realistic upper gas fraction limits have been calculated, assuming their
M/$L_B$ ratio to be 0.5, see text for further details.\\
(e) Values from Thuan, Izotov \& Lipovetsky\ (1996)
\end{minipage}
\end{table*}
Extreme upper limits have been obtained assuming that each of the galaxies are
experiencing their first burst, and are observed during maximum luminosity.
Indeed, this is an extreme upper limit, and should be regarded as such.
Adopting the evolutionary stellar models of Leitherer \& Heckman
\shortcite{leitherer:1995}, the $M/L_B$ ratio may be estimated for a burst, 
$M$ being the mass of stars.  Leitherer \& Heckman calculated the blue 
luminosity for a $10^6\,M_{\odot }$ starburst as a function of time. To be 
consistent, their Salpeter IMF has been scaled to ours. Indeed, two different 
IMF's have been used so far, but to be sure that upper gas fraction limits are 
obtained, the high lower limit, $m_{L}$=0.1$M_{\odot }$, is used, as this 
gives the lowest $M/L_B$, hence the highest gas fraction.
The scaling is very simple, since the only difference is to scale the lower
mass limit from 1 to 0.1 $M_{\odot }$. Hence, all luminosities from 
Leitherer \& Heckman are multiplied by 0.39. From their fig.\ 9 (Z=0.1 
$Z_{\odot }$), the minimum magnitude is read off to be \mbox{-16.4} 
corresponding to a maximum luminosity $\sim  5.6\times 10^8\, L_{\odot }$ 
using the absolute blue solar magnitude 5.48. After IMF scaling this gives
$\sim 2.2\times 10^8\, L_{\odot }$, so $M/L_B\sim 0.004$. In table\
\ref{gasfrac} the observed blue luminosity, scaled to h=0.65, is given for 
each galaxy. Table\ \ref{gasfrac} gives the corresponding gas fraction, as 
$\mu = M_{gas}/M_{gas}+M_{stars}$. If IZw18 is experiencing its first burst, 
this upper limit should be realistic as its true gas fraction. The large difference 
from the dynamically deriven gas fraction is caused by a combination of influence 
from DM, poor total dynamical mass estimation or/and that IZw18 is not 
experiencing its first burst.

To constrain the models more, three more galaxies have been included, for 
which we know that they are definitely not experiencing their first burst. One 
of them is the LMC. Strictly speaking, the LMC is not a dwarf galaxy, but in 
many ways it
behaves as such. For these three galaxies, $M/L_B$ is assumed to be 0.5, not as
extreme as above, thus obtaining upper limit gas fractions lower than for
the other galaxies. Does this assumption still provide an upper limit for the
gas fractions? The answer is confirmatory, if the assumed $M/L_B$ is a lower limit.
The ratio is estimated using the star formation histories of LMC 
\cite[fig.7(c)]{geha:1998}, 
SMC \cite{pagel:1998}, and NGC6822 \cite[fig.12]{gallart:1996}. The general 
procedure is quite simple. First we find $L_B$ of the objects by using a stellar 
population synthesis model and the SFH. The blue magnitudes in Charlot \& Bruzual 
\shortcite[fig.5, dashed line]{charlot:1991} are calibrated by using Leitherer \& 
Heckman \shortcite[fig.10, 0.25$Z_{\odot }$, solid line]{leitherer:1995}. When we know 
the age of a starformation epoch, we may find the blue magnitude, hence the luminosity 
by reading off figure 5 of Charlot \& Bruzual. The results are valid for 
$m_{L}$=1$M_{\odot }$, $m_{U}$=100$M_{\odot }$ and a starformation rate of 
1$M_{\odot }yr^{-1}$, since the magnitudes are calibrated by using Leitherer \& 
Heckman. By doing this we ignore the very few stars with masses between 100$M_{\odot }$
and 125$M_{\odot }$ originally included by Charlot \& Bruzual. This will not have any 
significant effect on the results.

For LMC we have the SFR in relative units. We have splitted the SFH into two star 
formation epochs, one that started 2 Gyr ago and one 12 Gyr ago. Integration of the 
SFR and weighting the two read-off luminosities with respect to SFR leads to 
$M/L_B$=0.4. By scaling the luminosities to $m_{L}$=0.1$M_{\odot }$ and 
$m_{L}$=0.01$M_{\odot }$, we obtain $M/L_B$=1.0 and 2.4, respectively.

For NGC 6822 we are given the SFR in absolute units, but we restrict ourselves to use
the relative SFR only. Again we split the SFH into two epochs. A very recent one that 
started 200 Myr ago and a very old one that started 15 Gyr ago. As we ignore the epoch
that stopped 5 Gyr ago, we will obtain a lower estimate of $M/L_B$. Using the method
outlined above, we get $M/L_B$=0.16, and 0.42, 0.96 for the 0.1 and 0.01 lower mass 
limits, respectively.

The lack of knowledge of the SFH in the SMC is remarkable. Hence, we have used the
SFH from the model by Pagel \& Tautvaisene \shortcite{pagel:1998}, which has 
succes in explaining e.g. [Fe/H]. The SFR is given in absolute units in their table 2,
so integration yields the total mass of stars. Using $L_B=0.99 \times 10^9\, 
M_{\odot }$ from table\ \ref{gasfrac}, we arrive at $M/L_B$=0.43.

In conclusion we have that $M/L_B$=0.5 is a conservative lower limit, hence giving
an upper limit on the gas fraction.

\begin{figure}
\epsfxsize=\linewidth
\epsfbox{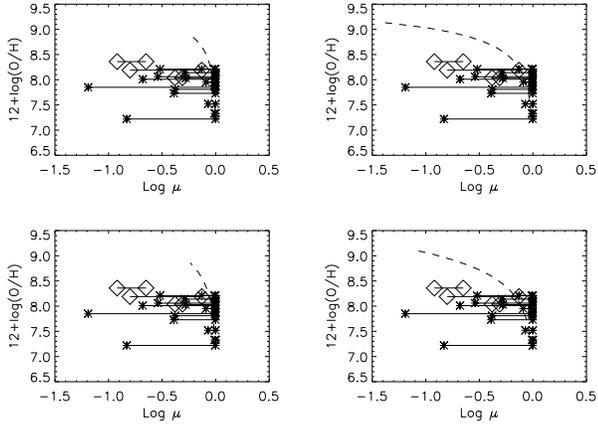}
\caption{12+log(O/H) as a function of gas fraction, 
$\mu \equiv \frac{M_{gas}}{M_{gas}+M_{stars}}$. Top plots are for set 2a, for
$m_L=0.1\: M_{\odot }$ to the left and $m_L=0.01\: M_{\odot }$ to the right. Bottom 
plots are for set 3, again $m_L=0.1\: M_{\odot }$ to the left and $M_{low}=0.01\: 
M_{\odot }$ to the right. The dashed lines are
the results of our closed models, using burst masses equal to $2\times 10^6\:
M_{\odot }$ for the plots on the left and $4\times 10^6\: M_{\odot }$ for the
plots on the right. The stars are the lower and upper limit gas fractions for 
the majority of the sample, connected by lines. The diamonds are the limits 
given for the three galaxies, having more moderate upper limits.} \label{closedgas}
\end{figure}
The results from the calculations are shown for yield set 2a (top plots) and 3 
(bottom plots) in fig.\ \ref{closedgas}. In both cases, using the IMF with
$m_L=0.1\: M_{\odot }$, the fits are always between the lower and upper limits, except 
for the three 'moderate upper limit' galaxies. Lowering the yield by adopting 
$m_L=0.01\: M_{\odot }$, helps somewhat, but only one of the three moderate galaxies 
is fitted, namely NGC 6822, the two others being out of range.

Since the model fails in explaining the more restrictively chosen
galaxies, the yields have to be lowered further. The number of low-mass stars 
could be increased, but it would be parameter-gambling, not appealing very
much to a physical control of the model. However, it should be noted that Carigi 
et al.\shortcite{carigi:1998} use this possibility to solve the problem. 
Keeping the IMF, the only way of 
lowering the yield is to open the model and allow for galactic winds, ordinary 
or enriched \cite{carigi:1995} or/and inflow \cite{pagel:1998}.  

\subsection{The model including enriched winds}
From this place forth, the model is opened, allowing for gas exchanges with
the intergalactic medium. All simulations have used $m_L=0.01\, M_{\odot}$.

The results of including enriched winds are shown in fig.\ \ref{richRV} for set 2a 
and fig.\ \ref{richpad} for set 3. 
As seen, the wind efficiencies employed are the lowest acceptable for fitting
the gas fractions of the Magellanic clouds. Using the same model, the 
outputs have been plotted with the observations for both N/O-O/H and Y-O/H. 

However, removing that much oxygen raises
the N/O ratio high above the level of the observations. One could try to
reduce N/O by lowering the value of $\alpha $, in the case of set 2a, but a
value much lower than 1, seems to be quite unrealistic, seen in the light of
recent work on stellar evolution, e.g. Marigo \shortcite{marigo:1998}.

For set 3 (fig.\ \ref{richpad}), it is clearly seen that setting the 
primordial He fraction equal to its lower uncertainty value, as done when using 
set 3 in the closed model (see fig.\ \ref{ypad}), is not capable
of explaining the He fractions, when introducing these high-efficient
enriched winds - the fits are much too poor.

Hence, introducing enriched winds makes it impossible
to fit gas fractions and N/O simultaneously. Further, in the case of set 3,
problems also arise in fitting the He fractions.
\noindent
\begin{figure}
\epsfxsize=5cm
\epsfbox{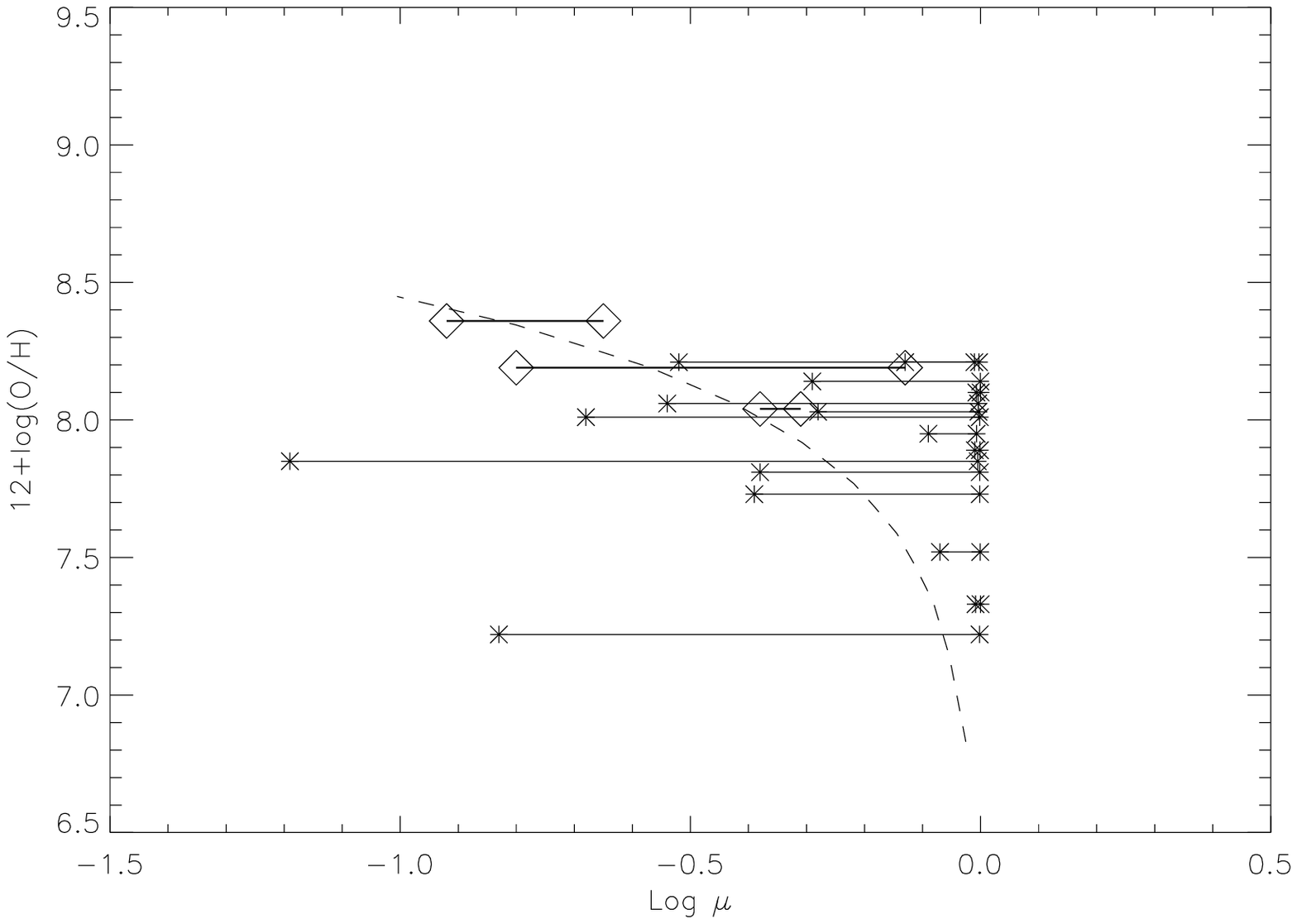} \vfill
\epsfxsize=5cm
\epsfbox{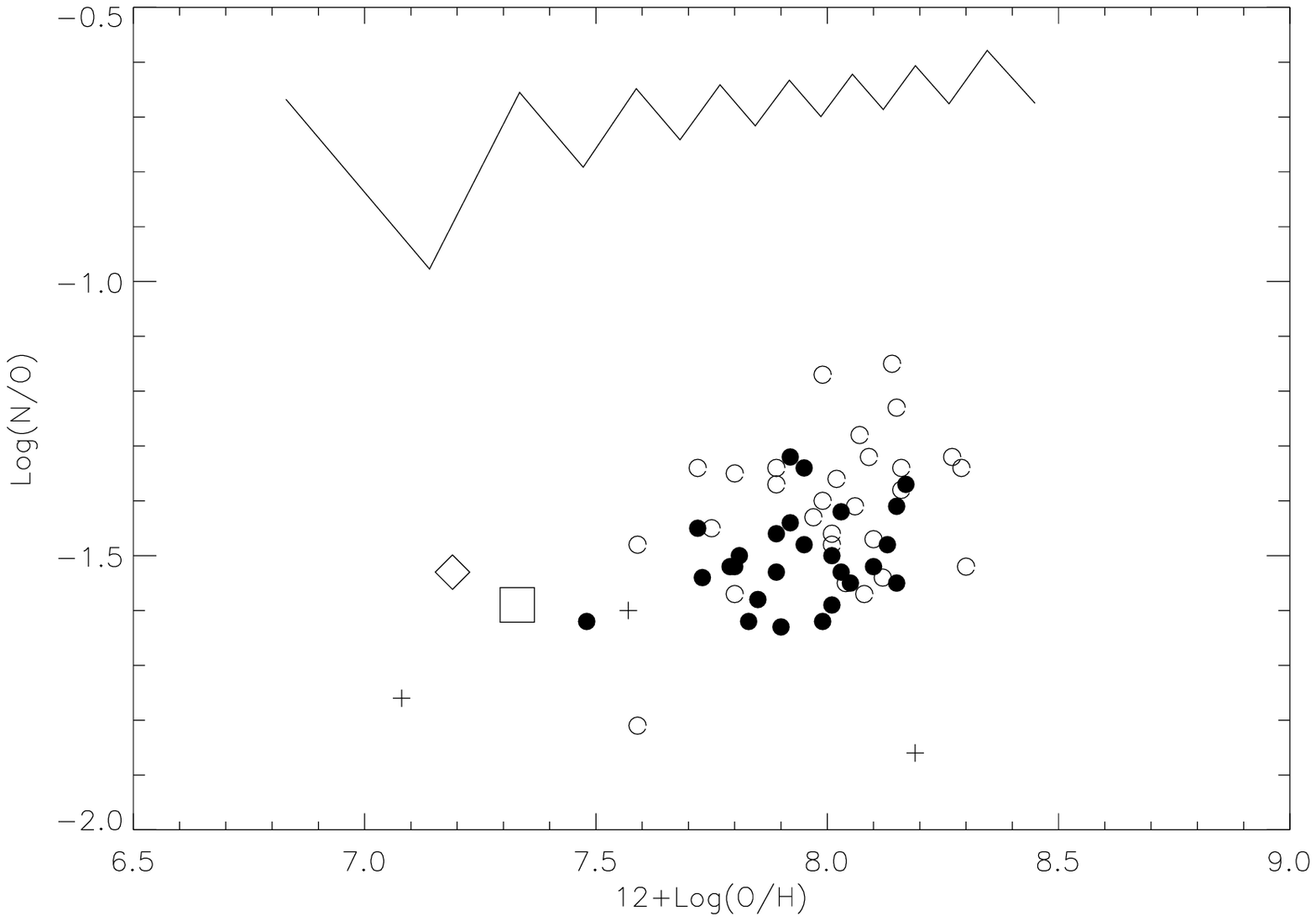} \vfill
\epsfxsize=5cm
\epsfbox{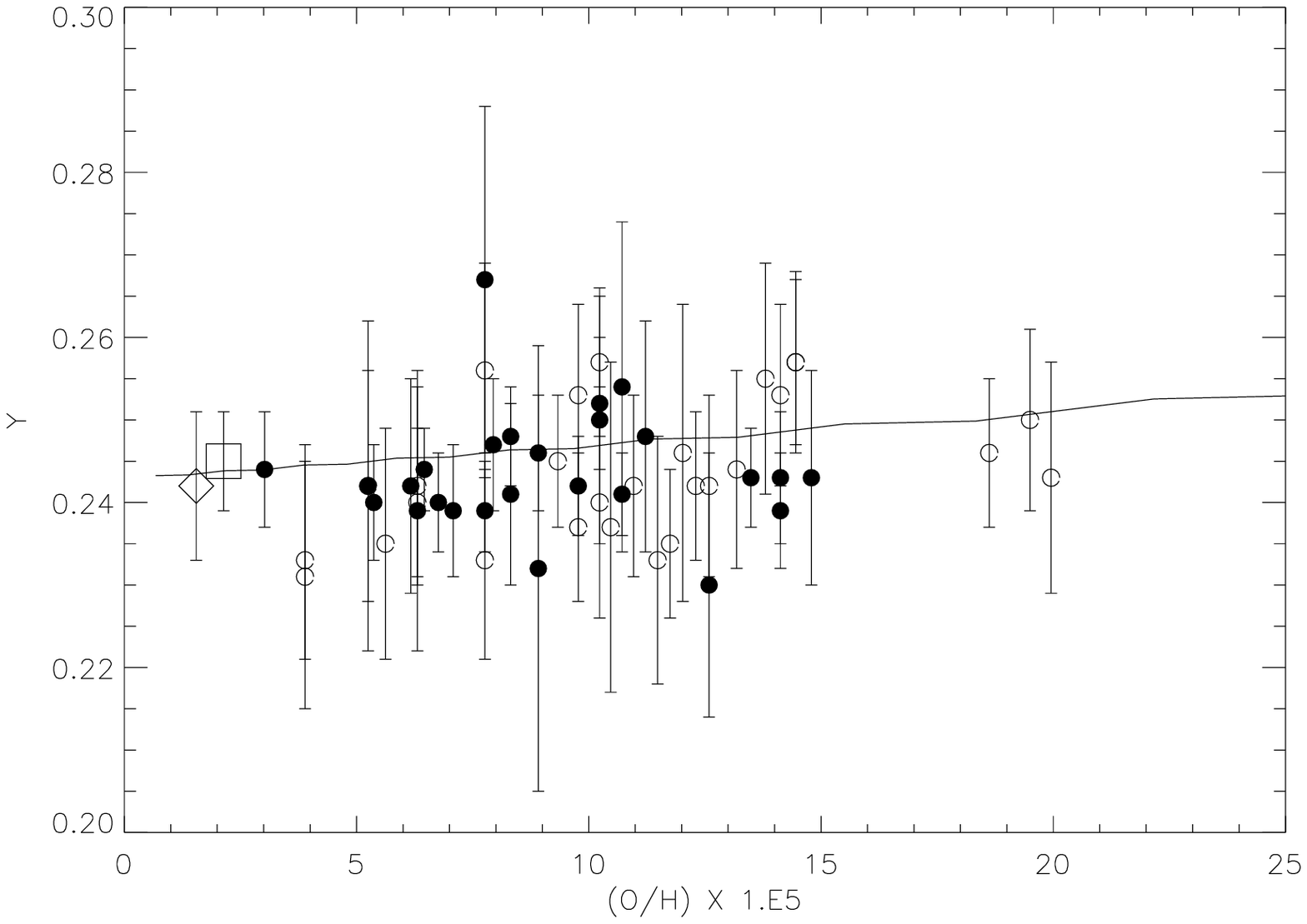} 
\caption{The results of models with enriched winds. The plots are made, 
using an efficiency factor equal to 0.8, adequately to fit the gas fractions. The 
yield set in use is 2a. The values of
$M_{HBB}$ and $\alpha $ are the same as used for the closed models, thus being
5 $M_{\odot }$ and 1.1, respectively. All burst masses are equal to
$6\times 10^6\: M_{\odot }$.}
\label{richRV}
\end{figure}

\begin{figure}
\epsfxsize=5cm
\epsfbox{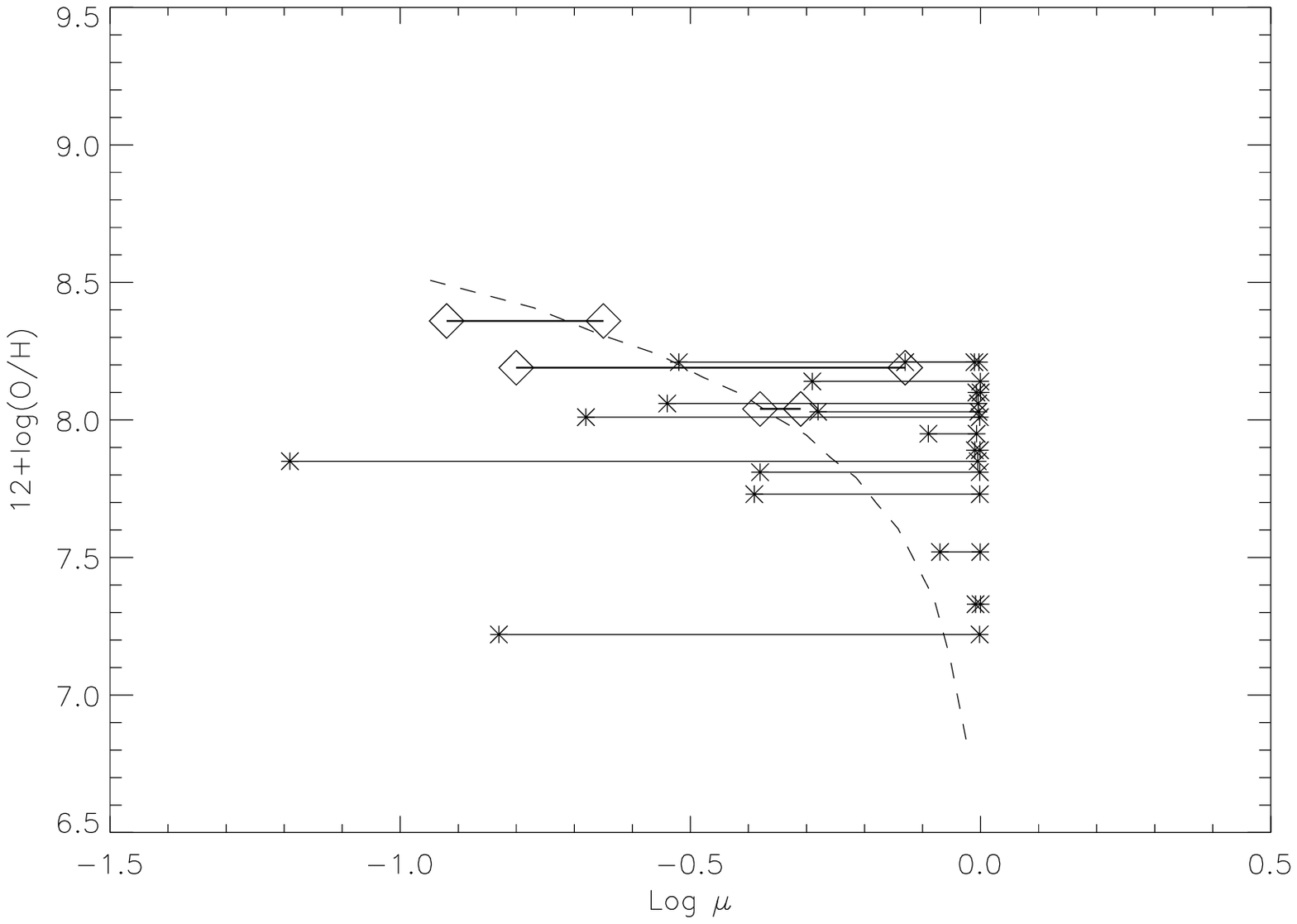} \vfill
\epsfxsize=5cm
\epsfbox{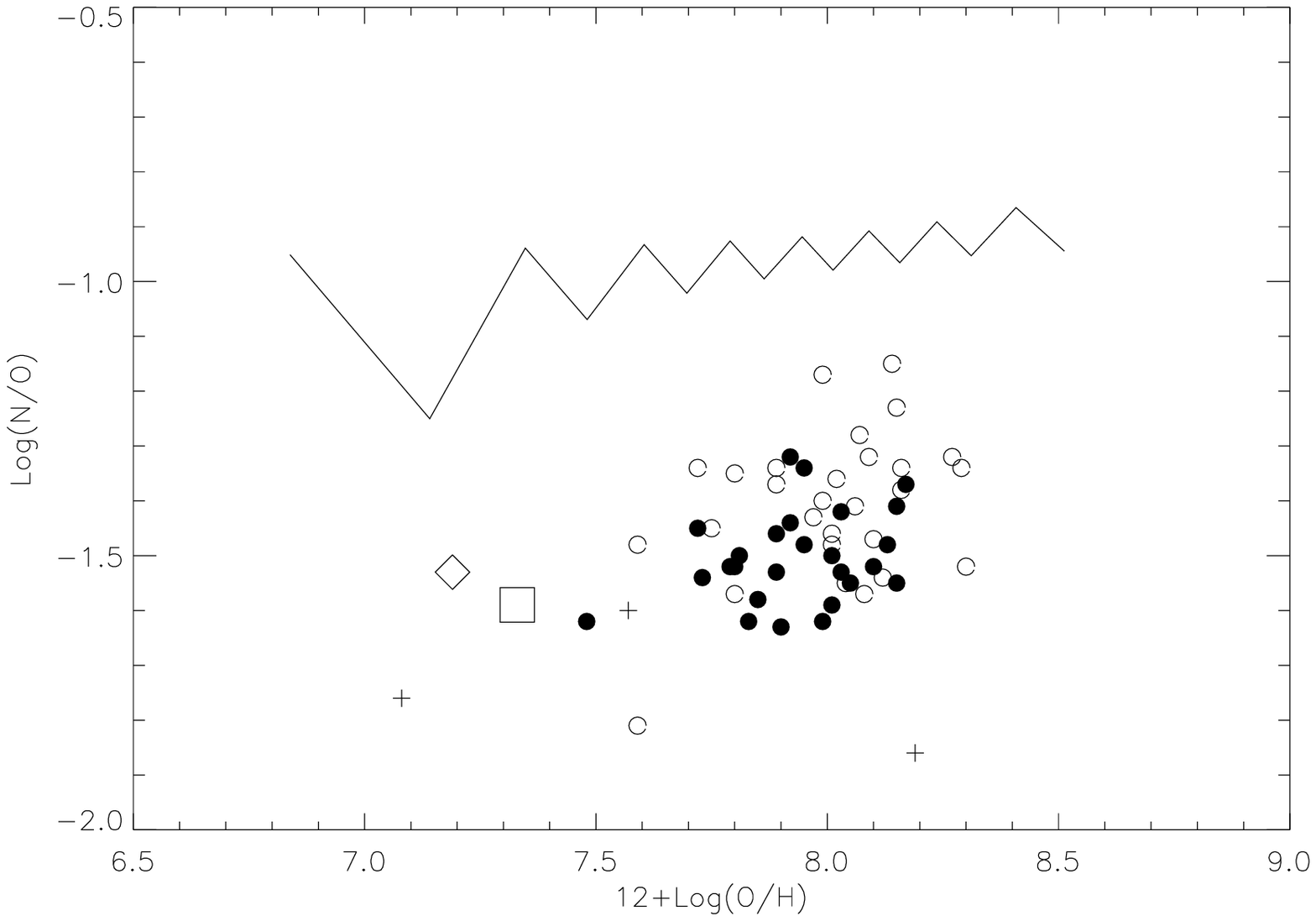} \vfill
\epsfxsize=5cm
\epsfbox{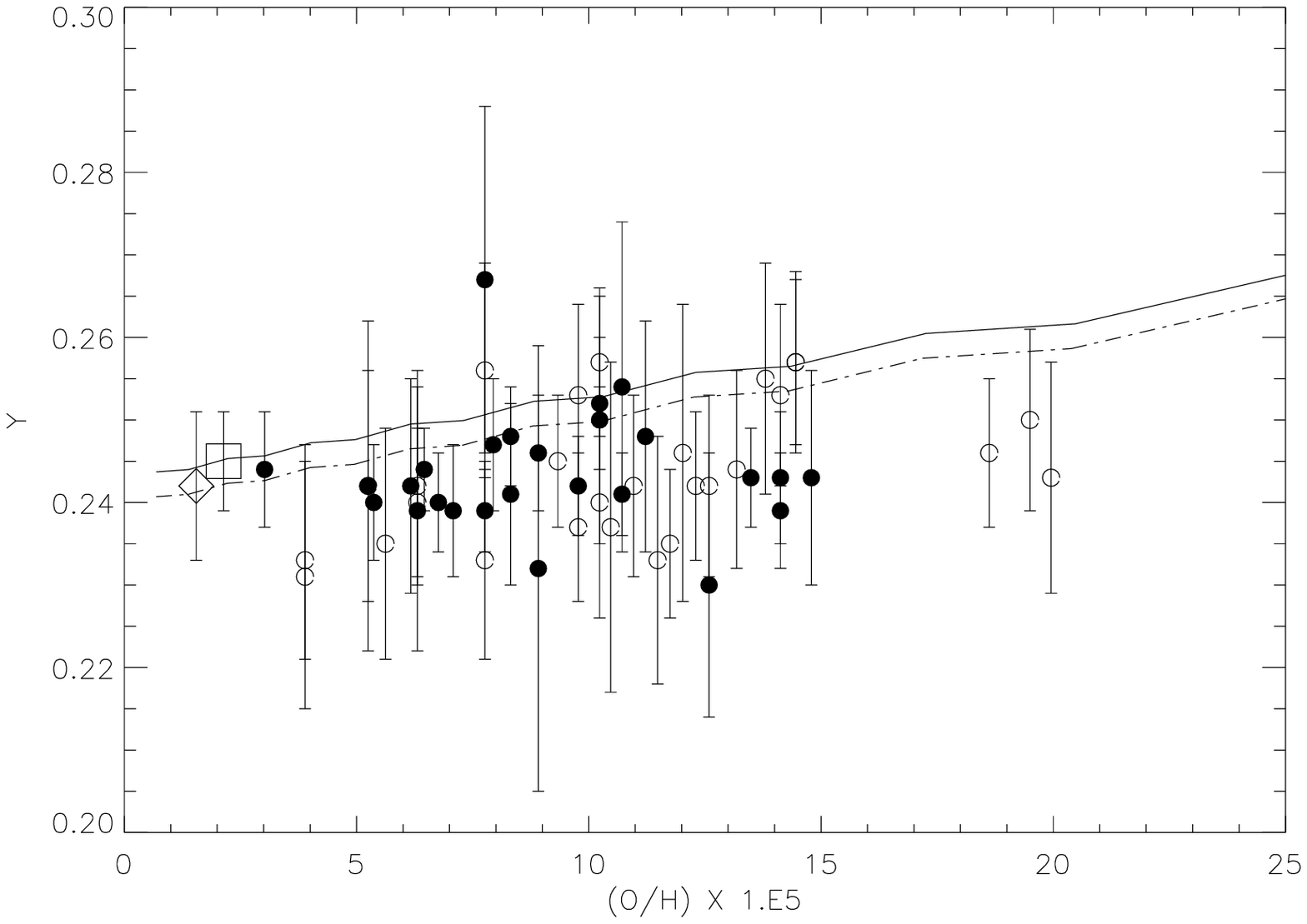}
\caption{The results using the same parameters as in fig.\
\ref{richRV}, but with yield set no. 3 (Padova yields). The efficiency
factor is 0.7, chosen to fit the gas fractions.}
\label{richpad}
\end{figure}

\subsection{The model including ordinary winds}
Though keeping the mass of each burst as a free parameter, it turns out that
$2\times 10^6\, M_{\odot }$ is fulfilling the requirement on placing the
second burst close to the abundances of IZw18. Thus, it is appropriate to keep 
this burst mass constant in the following discussion.
The results are shown, in fig.\ \ref{ordinaryRV} for set 2a and
fig.\ \ref{ordinarypad} for set 3.

It was shown above that the scatter in N/O could be explained by our model.
Hence, the scatter will not be given much attention 
in the following, as the work will be concentrated on getting the right level
of N/O, coincident with the gas fraction fitting, if possible.

From the top plots, it is evident that a mass removal of 5 times the burst
mass does not appreciably change the appearance of the fits from the closed
model. Actually, the O-yield is lowered only by
$\sim $0.3 dex as seen when comparing to fig.\ \ref{closedgas}, using 2a
yields. The point may be that we are dealing with
a discontinuous star formation. Note for instance the upper linear part of the
model curve. The increase in O/H along this part is due to one burst
only, as seen when comparing to the corresponding N/O plots. The
explanation is that the smaller the absolute gas mass, the larger the effect
of oxygen enrichment. Thus, even when adopting higher wind efficiencies, it
is not possible to bring the oxygen yield sufficiently down.
The same mechanism is responsible for increasing the scatter in N/O for
higher wind efficiencies. Hence, it is possible to produce large scatter
without having a large burst mass.
 
Note also that the evolutionary tracks stay close to a gas fraction of 1 for
a relatively large part of the evolution. A simple calculation will show, how
difficult
it is to bring the gas fraction down. For instance, assume the first burst to
turn $2\times 10^6\: M_{\odot }$ into stars, and set $W_{ISM}=5$. Then, just
before the second burst, the gas fraction will be
\BA
\mu \equiv \frac{M_{gas}}{M_{tot}} &=& \frac{M_{gas}(0)-2\times 10^6\:
M_{\odot } - 10^7 M_{\odot }}{M_{tot}(0)-10^7 M_{\odot }} \\
 &=& 0.98, \nonumber
\EA
\noindent
since the initial mass of the dwarf galaxy is $M_{gas}(0)=M_{tot}(0)
=10^8\, M_{\odot }$. Using the equation again, one finds $\mu =0.95$ just
before the third burst. These values are lower limits (!), since the increase 
in gas from stellar ejecta is disregarded.

Both the level of N/O and slope of Y
are satisfied. Note the similarity between the Y plots presented here and
those presented by the closed model. This is not surprising, as about half of
the helium is produced by massive stars, hence following oxygen.
Status is that all observations are matched reasonably well by including ordinary
winds, but no better than for the closed model. In particular, it is found that
it is impossible to bring the yields sufficiently down to explain the gas
fraction intervals of the Magellanic clouds.
\begin{figure}
\epsfxsize=5cm
\epsfbox{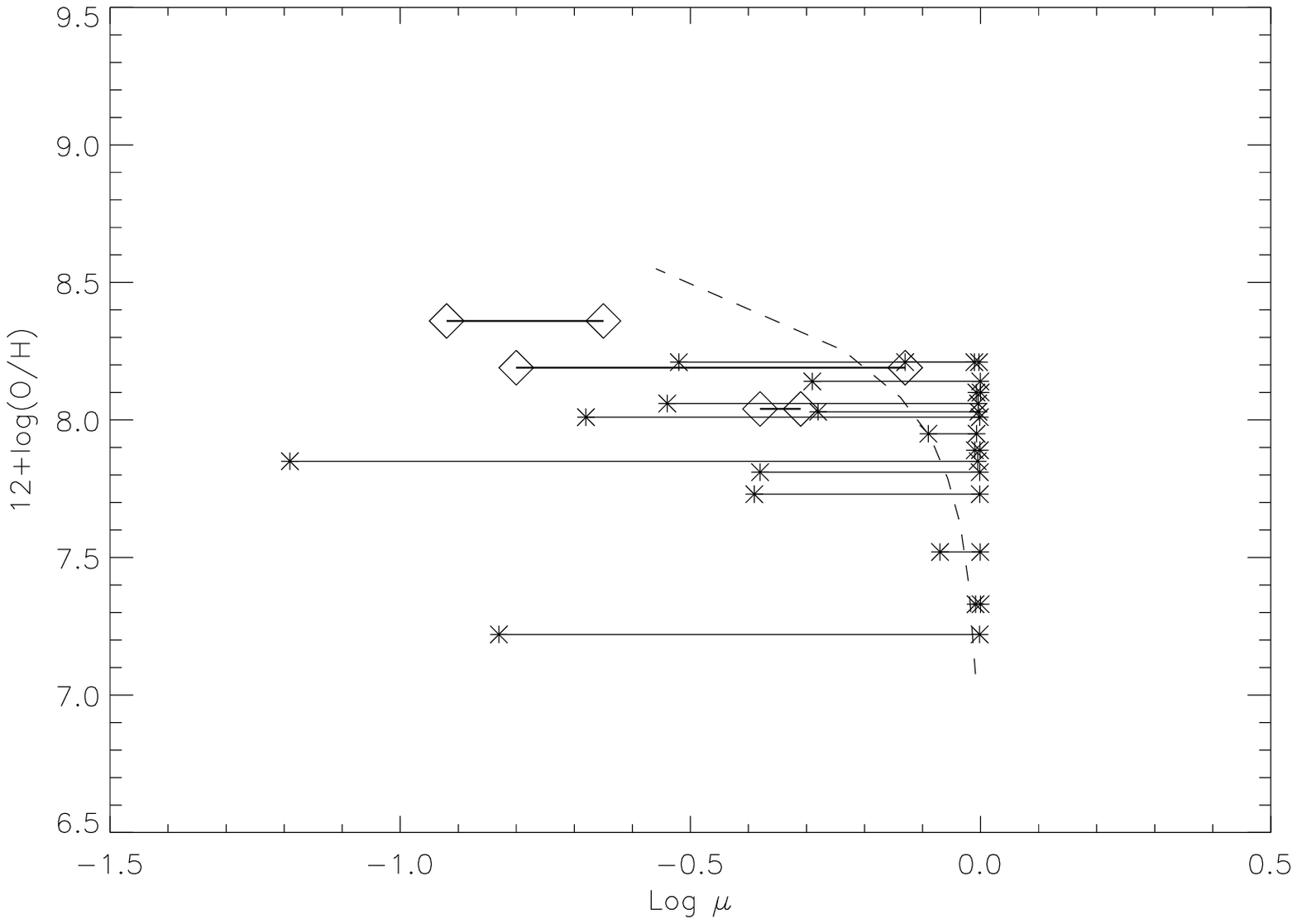} \vfill
\epsfxsize=5cm
\epsfbox{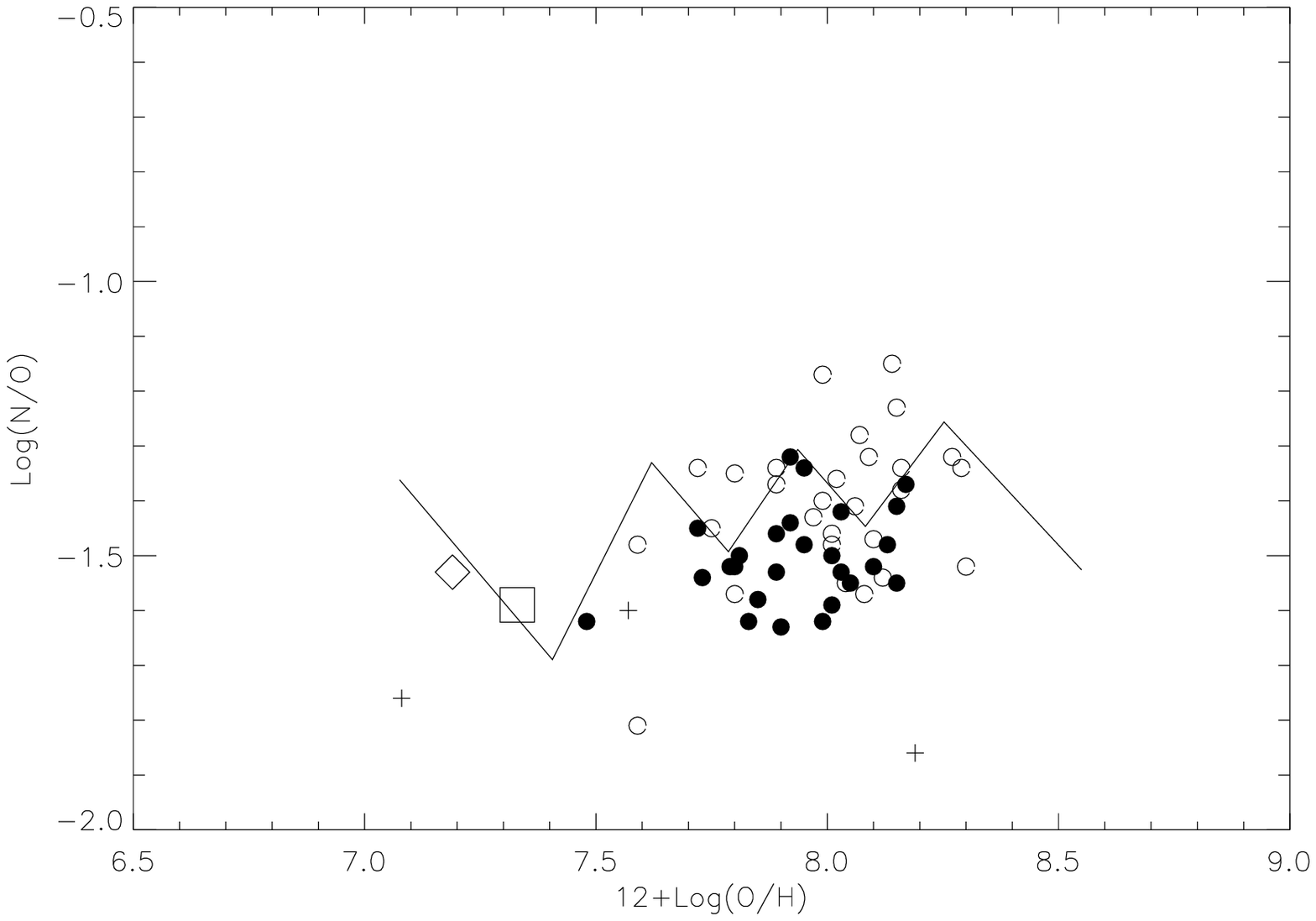} \vfill
\epsfxsize=5cm
\epsfbox{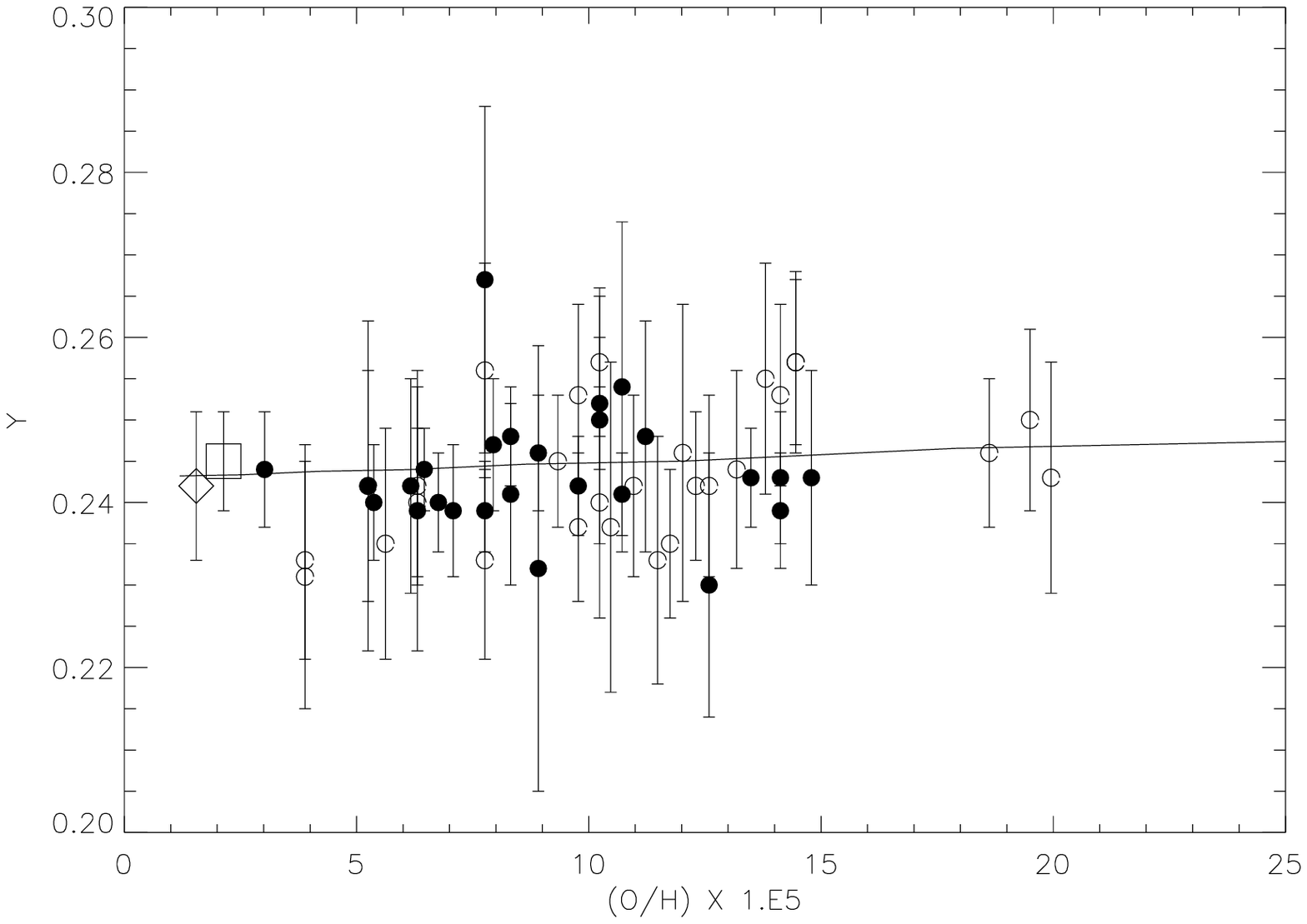}
\caption{The results of using a model including ordinary
winds. The plots are made using $W_{ISM}=5$.
Note the increasing scatter in N/O as a function of $W_{ISM}$ explained in the
text. The yield set in use is 2a. The values of
$M_{HBB}$ and $\alpha $ are still 5 $M_{\odot }$ and 1.1, respectively, and the
mass of each burst is $2\times 10^6 M_{\odot }$.} \label{ordinaryRV}
\end{figure}
 
\begin{figure}
\epsfxsize=5cm
\epsfbox{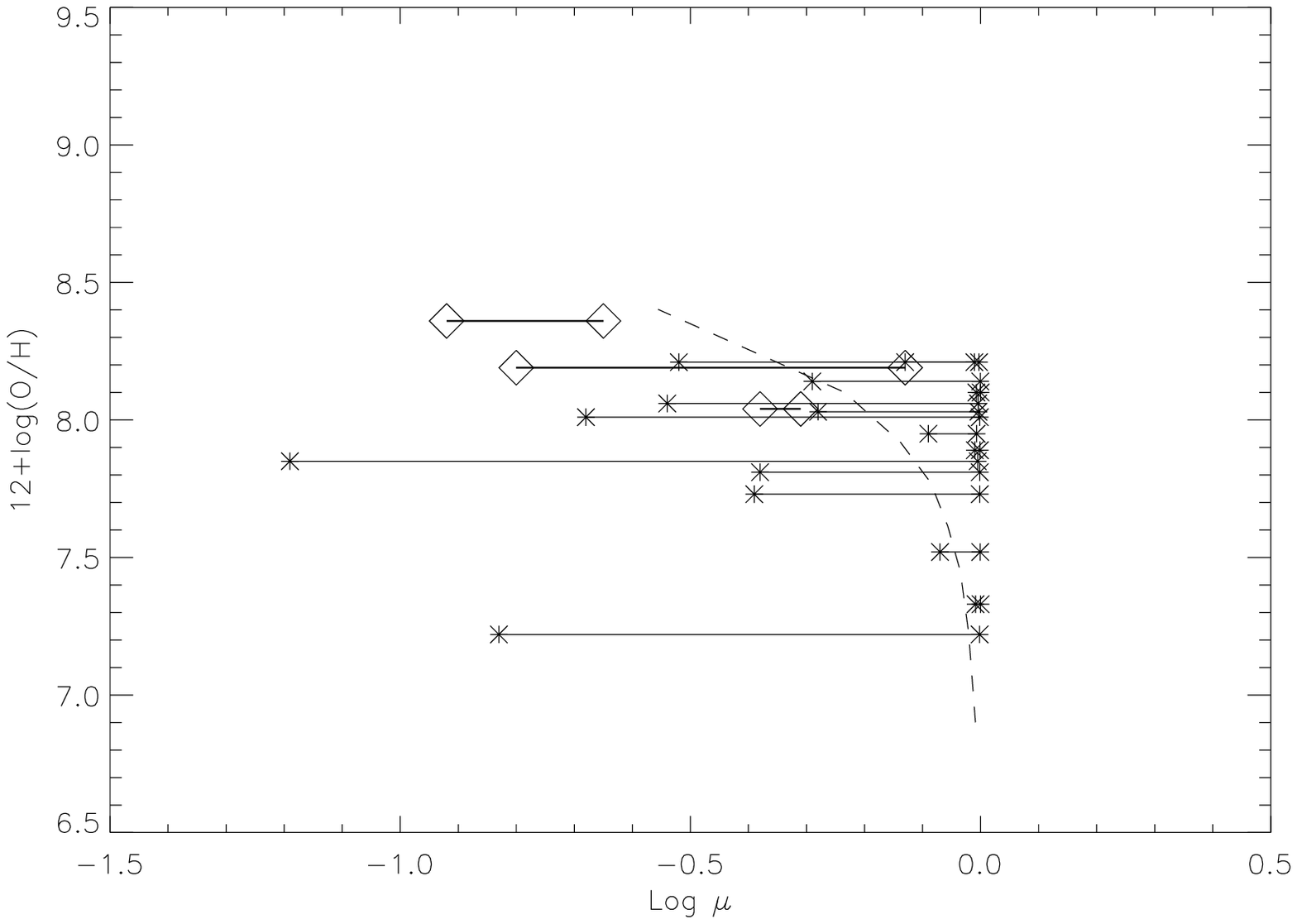} \vfill
\epsfxsize=5cm
\epsfbox{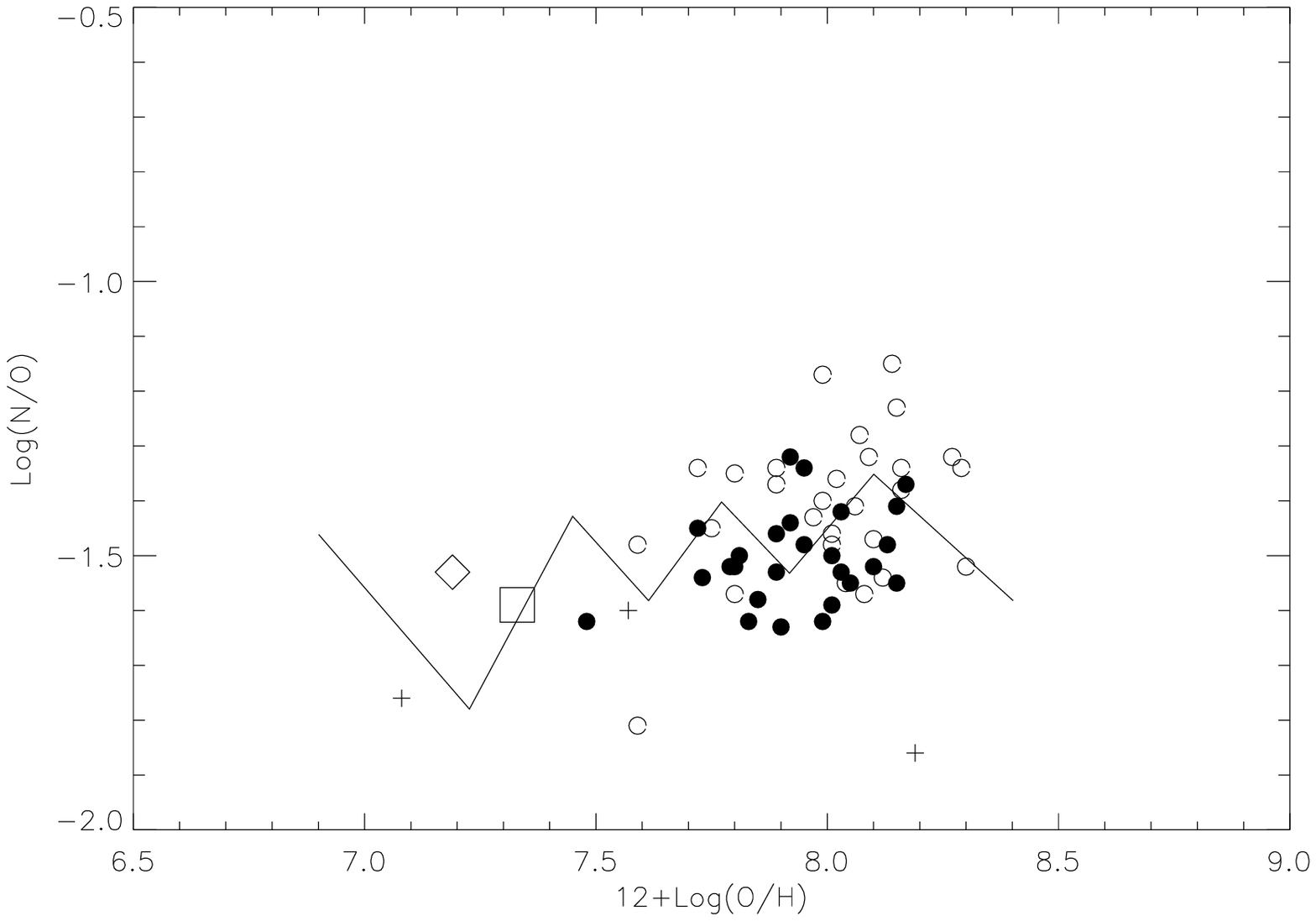} \vfill
\epsfxsize=5cm
\epsfbox{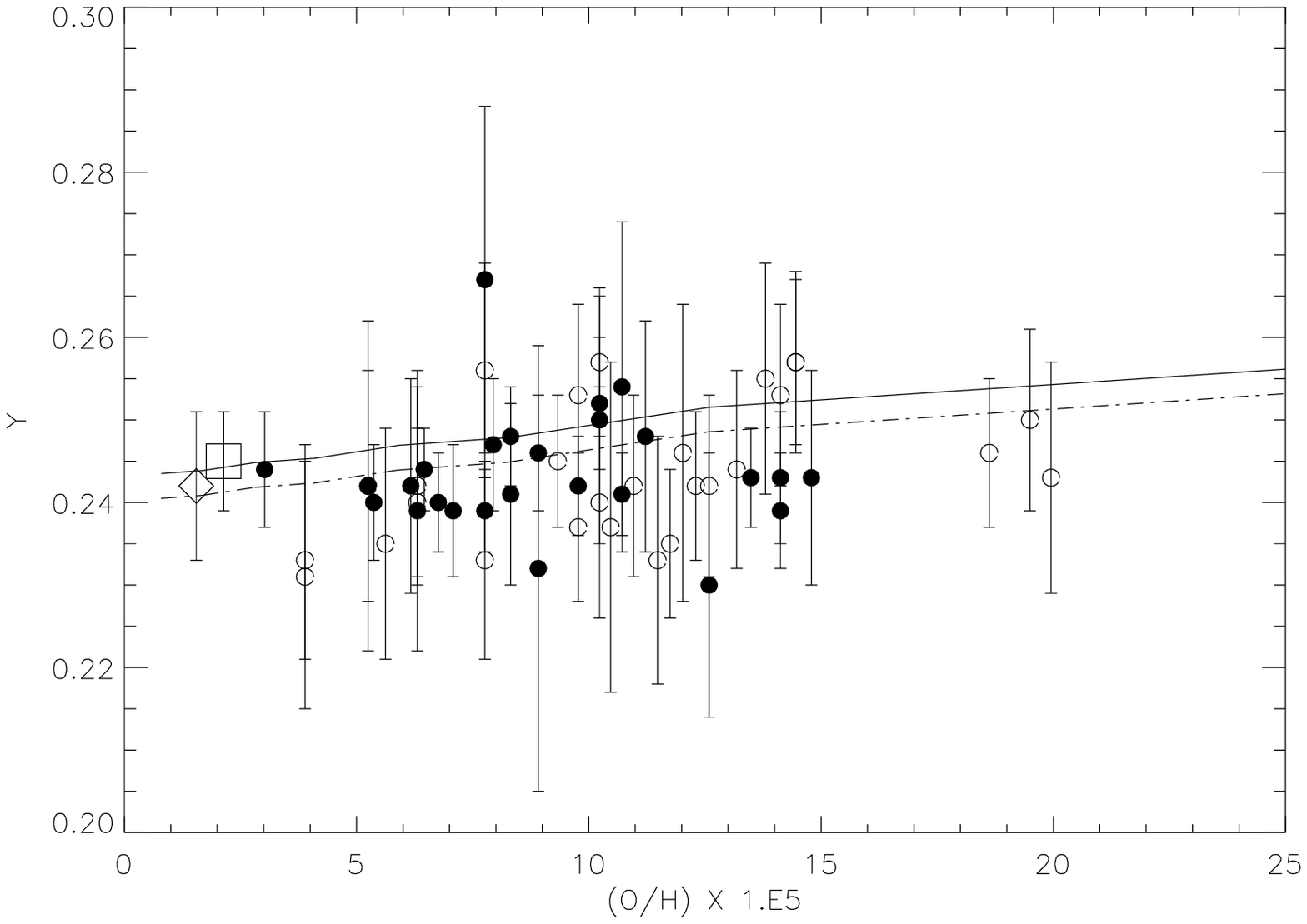}
\caption{The results using the same wind and mass parameters as
in fig.\ \ref{ordinaryRV}, but with yield set no. 3.} \label{ordinarypad}
\end{figure}

\subsection{Comparing the ordinary wind model with an analytical model}
A few calculations have been performed, using a simple analytical model, 
employing continuous star formation and Hartwick-outflow.
The outputs of the analytical model and the numerical model are then compared.
The comparison is only performed for yield set 2a, since
the inclusion of winds is identical using set no. 3.
 
The general formulation of Hartwick-outflow is
\BA
\frac{dM_{dw}}{dt}=-W_{ISM}\frac{dM_s}{dt}
\EA
 
\noindent where $M_{dw}$ is the mass of the dwarf galaxy and $M_s$ is the
mass of stars. The next step is to use
 
\BA
\frac{dZ}{dM_s}=\frac{p}{M_g}
\EA
 
\noindent obtained by considering the changes in Z as a result of stellar
ejecta, star formation and outflow, see e.g. Pagel
\shortcite[his eq. 7.37]{pagel:1997}. Instantaneous recycling is
assumed, and so is the absence of inflow. Separating the variables and using
\mbox{$\delta M_g=-\delta M_s-W_{ISM}\delta M_s$}, one finds
\begin{figure}
\epsfxsize=\linewidth
\epsfbox{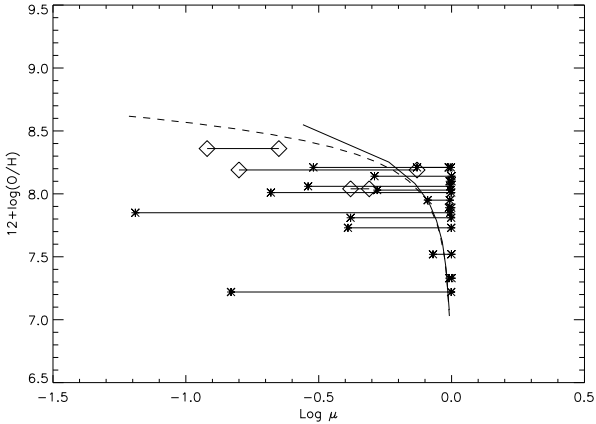}
\caption{The comparison between the numerical bursting model
employing ordinary winds with $W_{ISM}$=5 and burst masses $2\times 10^6\:
M_{\odot }$, and the continuous star forming analytical model. The
solid line is the numerical bursting model.} \label{comp5}
\end{figure}
\begin{figure}
\epsfxsize=\linewidth
\epsfbox{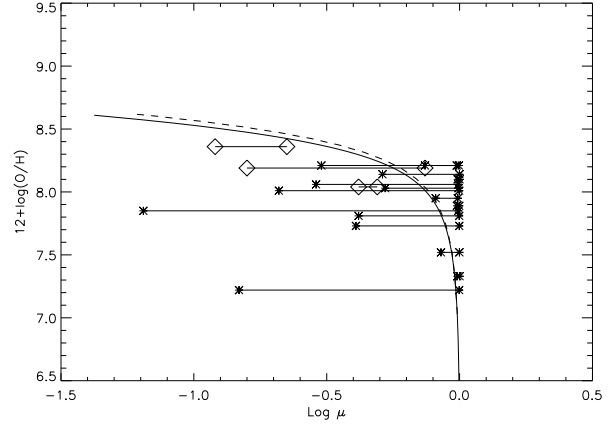}
\caption{Comparing the analytical model from fig.\ \ref{comp5}
with the  semi-continuous numerical model. The solid line is the numerical 
model and the dashed line is the analytical model. See text for details.}
\label{compcont}
\end{figure}
\BA \label{open}
 \frac{Z}{p}= \frac{1}{1+W_{ISM}}\ln \left( \frac{M_g(Z=0)}{M_g}\right)
\EA
\noindent 
where the yield is assumed constant. The total mass of the dwarf
galaxy is written
 
\BA
M_{dw} &=& M_{dw}(Z=0)-W_{ISM}M_s \\
    &=& M_g(Z=0)-W_{ISM}(M_{dw}-M_g)
\EA
 
\noindent Isolating $M_g(Z=0)$ and inserting it into eq.\ \ref{open} gives
\BA \label{gasfraction}
\mu &=& \frac{1+W_{ISM}}{\exp \left( \frac{Z}{p}(1+W_{ISM})\right) +W_{ISM}} \\
    &=& \exp \left( -\frac{Z}{p}\right) \frac{1+W_{ISM}}{\exp \left(
 W_{ISM}\frac{Z}{p}\right) +W_{ISM}\exp \left( -\frac{Z}{p}\right)} \nonumber
\EA
 
\noindent where $\mu $ is the gas fraction $\frac{M_g}{M_{dw}}$. It is seen
that for a closed model, $\mu =\exp \left( -\frac{Z}{p}\right)$. For our
numerical model, the yield is metallicity-dependent. Hence, 
$\exp \left( -\frac{Z}{p}\right)$, equal to the gas fraction at Z for a
characteristic value of p, is found from the closed numerical model.
Inserting this into eq.\ \ref{gasfraction}, finally gives the gas fraction of
the analytical model, now employing metallicity-dependent yields and outflow.
 
The results are shown in fig.\ \ref{comp5} for $W_{ISM}$=5. The numerical fit 
(solid line) is the same as in fig.\ \ref{ordinaryRV}. Note that for a large
part of the evolution, the two models are coincident, but at the last part they 
are differing from each other. It is
likely that the difference is a result of using instantaneous bursts instead of
a continuous SFR. The point may be that the mass of gas is small at the last
part of the evolution. Thus, the relative amount of ejected oxygen from
one burst, is high compared to the gas mass, hence giving a large increase in
O/H.
 
To make a further check on this hypothesis, a continuous version of the 
numerical model has been made. This is done in an approximate way, introducing 
the following changes to the above numerical outflow model:
\begin{enumerate}
\item The bursts are all single bursts, i.e. all interburst periods are the
same.
\item The interburst periods are made short. The shorter the interburst
period, the more continuous the star formation. An interburst period of 3 Myr
was sufficient for our purpose.
\item The masses of the bursts are calculated by assuming a mean SFR. The
calculations here assume $<\rmn{SFR}>$=0.002 $M_{\odot }\, yr^{-1}$, 
corresponding to one burst of mass
$2\times 10^6\: M_{\odot }$ every Gyr. A reasonable value compared to the
calculations above. The masses of the bursts are then equal to 0.002
$M_{\odot }\, yr^{-1} \cdot $ 3 Myr = 6000 $M_{\odot }$.
\end{enumerate}

All other calculations are performed exactly the same way, as they were, using
the double bursting model.
The calculations are performed for $W_{ISM}$=5 and compared to the
analytical model from fig.\ \ref{comp5}. From fig.\ \ref{compcont}, it is seen 
that the two models are almost identical. The small offset may be caused by the 
calculation of the analytical model, adopting the yields from the closed 
numerical bursting model.
The similarity between the two models confirms the suspicion that instantaneous
bursting models behave differently from continuous models, as the numerical
model is the same that produced the fit in fig.\ \ref{comp5}, except for
the semi-continuous star formation.

\subsection{The model including ordinary winds and inflow}
The introduction of inflow gives two new parameters: the total mass accreted 
onto the galaxy $M_0$ and the duration of the inflow event $\tau _{inf}$. Only 
results for a few parameter choices are shown and interpreted by close 
examination of the results.
 
The gas fractions of the three selected
galaxies provide strong restrictions on the possible parameter space,
since inflow and outflow balance each other for obtaining the right gas
fraction values. Hence, if one increases the rate of inflow, one has to
increase the rate of outflow as well. Otherwise, the modelled gas fractions
would be too high, and the three selected gas fraction intervals would never
be reached.

It is assumed that the mass of
the original gas cloud is 0 at t=0. However, inflow of gas increases the
mass of the gas cloud, until it starts to form stars at a mass threshold
$M(t')$. The time $t'$ is calculated using eq.\ \ref{infmodel}, giving
\BA
t'=\tau _{inf}\ln \left( \frac{M_0}{M_0-M(t')}\right)
\EA
\noindent
For consistency with previous calculations, the threshold is assumed to be 
$10^8\: M_{\odot }$. Thus, after a time $t'$, the first burst appears. As for 
all the previous models, the first burst is a single one, and all other 
bursts appear in pairs. Hence, the appearance of the first burst is delayed by 
the time $t'$, when comparing to the models not including inflow. This time 
difference is included, when calculating the age of the system. Remember that 
the model terminates if the system is older than 15 Gyr.
\begin{figure}
\epsfxsize=5cm
\epsfbox{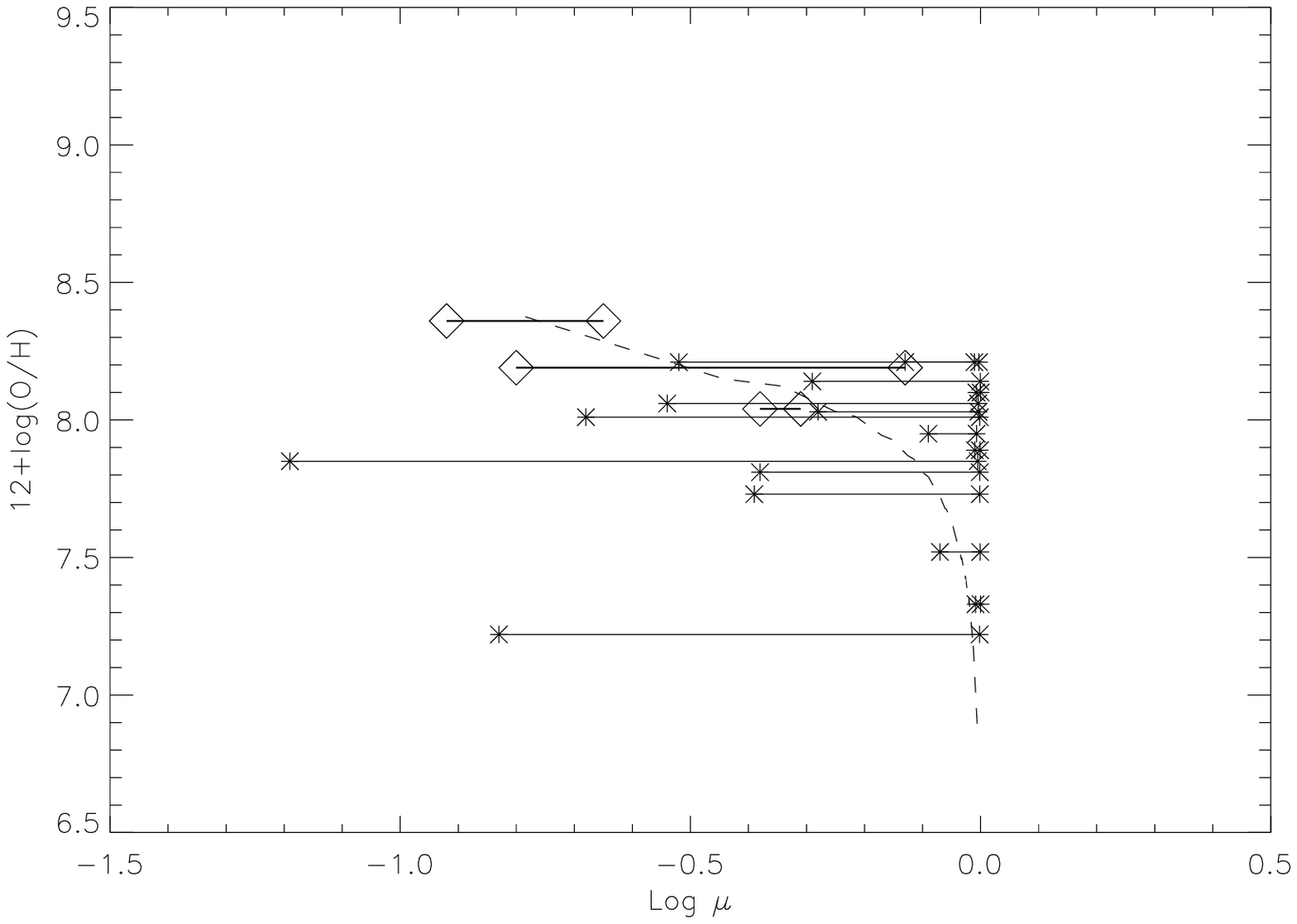} \vfill
\epsfxsize=5cm
\epsfbox{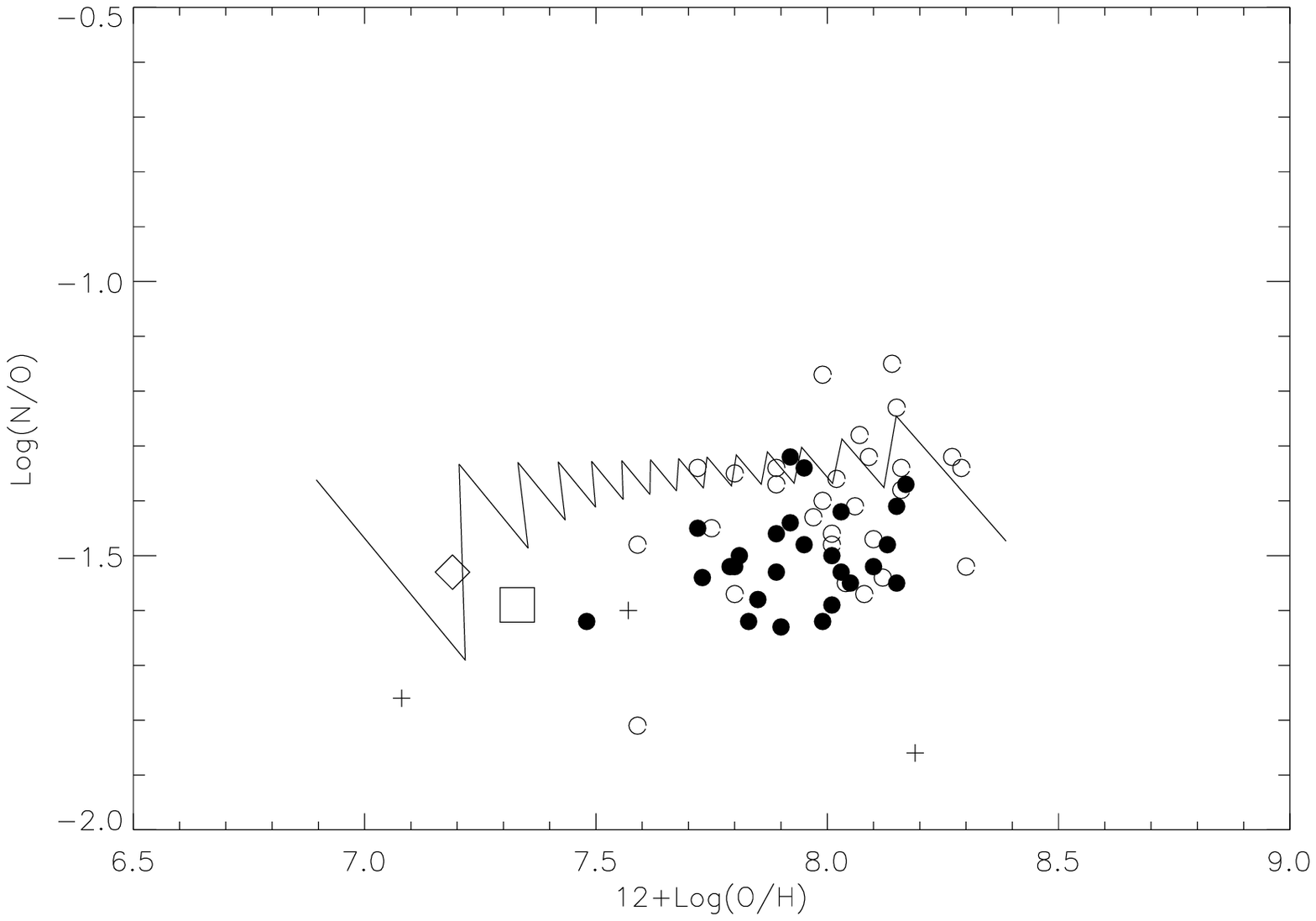} \vfill
\epsfxsize=5cm
\epsfbox{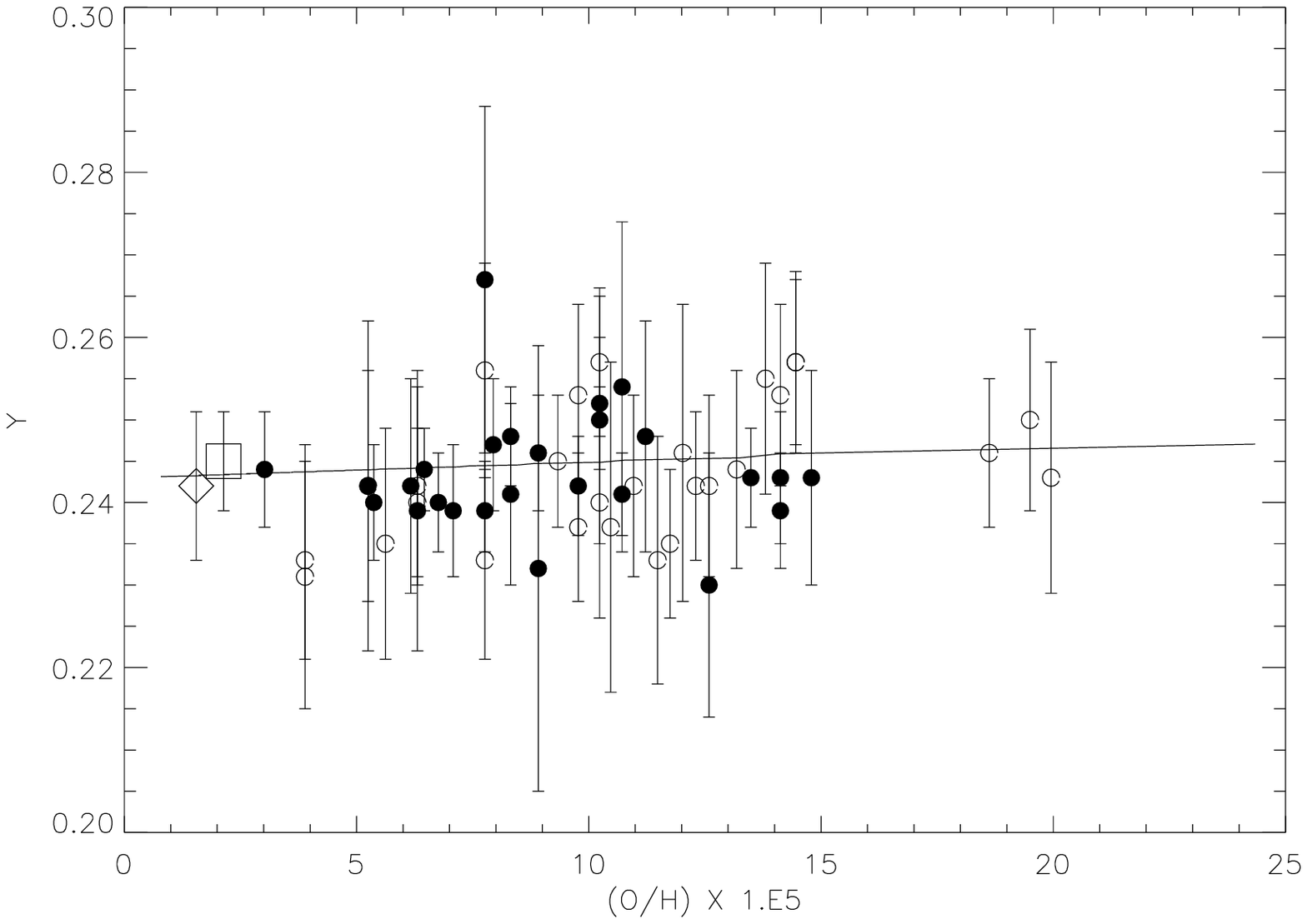}
\caption{The results of including both inflow and ordinary winds. Here the 
yield set used is no. 2a. The plots show a run using $\tau _{inf}$=5 Gyr
and $M_0=8\times 10^8\: M_{\odot }$. The burst masses are $3\times 10^6\: 
M_{\odot }$, and $W_{ISM}$=8.} \label{infRV}
\end{figure}

\begin{figure}
\epsfxsize=5cm
\epsfbox{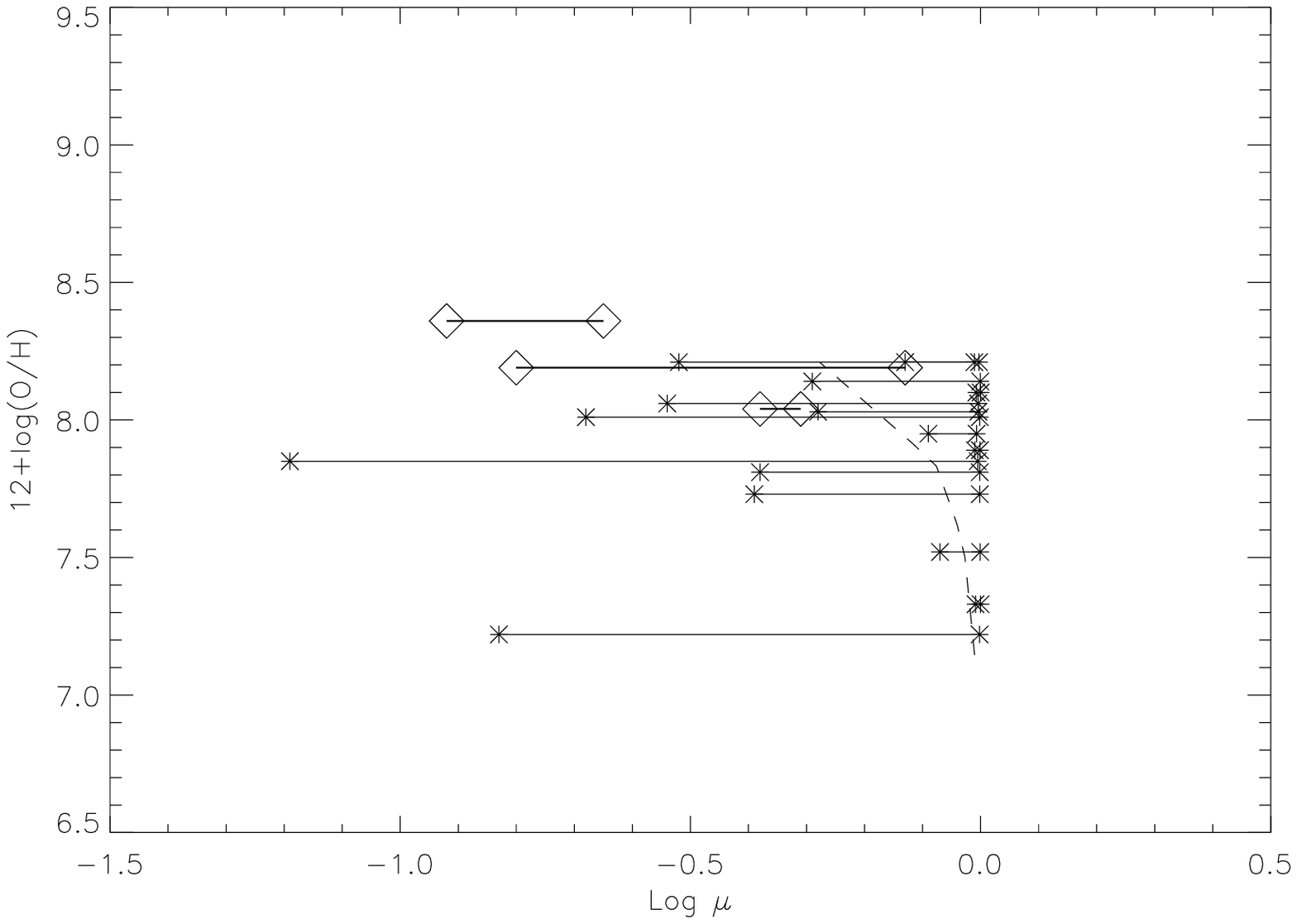} \vfill
\epsfxsize=5cm
\epsfbox{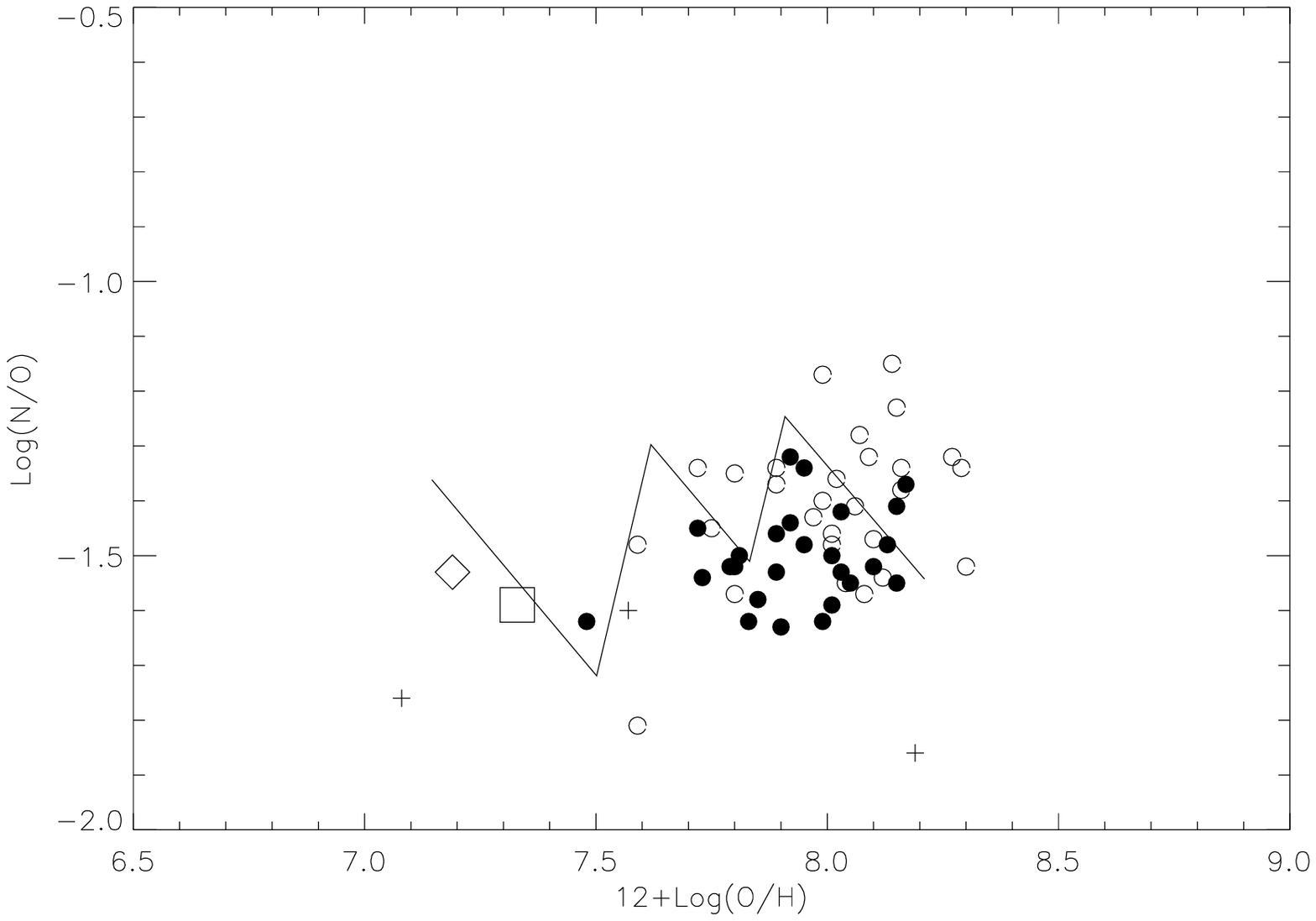} \vfill
\epsfxsize=5cm
\epsfbox{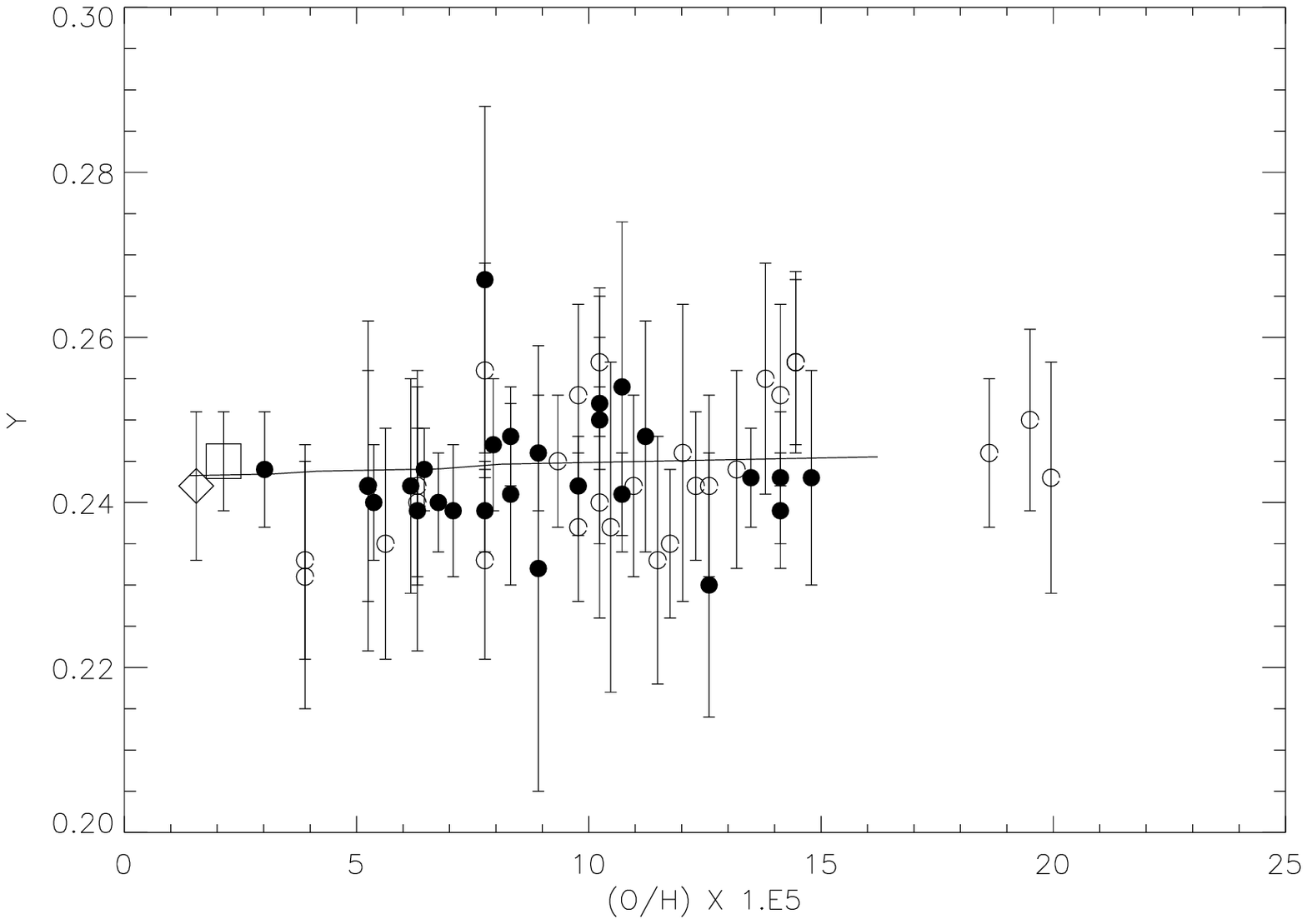}
\caption{Still using set 2a, though trying to increase the degree of scatter.
Now, the two parameters of inflow are $\tau _{inf}$= 2 Gyr and $M_0=2\times 
10^8\: M_{\odot }$. The burst masses and wind parameter have the same values as
in fig.\ \ref{infRV}.} \label{infRV2}
\end{figure}
 
For yield set 2a, the results are displayed in figs.\ \ref{infRV}, 
\ref{infRV2} and figs.\
\ref{gaslargescatter}, \ref{NOlargescatter}.

For the parameter values chosen very small scatter results
in the N/O diagram in fig.\ \ref{infRV}. The plots were made using a high
inflow/outflow rate to lower the yield sufficiently to fit the gas
fractions of NGC 6822 and the Magellanic clouds. The price to pay seems to be
a very small scatter in N/O. Hence, in fig.\ \ref{infRV2} a low rate of inflow
has been assumed, in particular at late stages of evolution, as $\tau _{inf}$
is decreased to 2 Gyr. Now, it is possible to explain the scatter in N/O, but
the result is an unsatisfactory gas fraction fit. Obviously, a large inflow
rate of primordial gas buffers the impact of O-rich ejecta on the ISM
abundances, resulting in small scatter.
 
To investigate further, whether it is possible or not to explain the
scatter in the N/O-O/H diagram and gas fractions simultaneously,
the outflow parameter is kept at 8, but the burst masses are increased to
$8\times 10^6\: M_{\odot }$ to give a larger scatter. To balance the high
outflow rate (remember that the outflow rate is proportional to the burst
mass) a rather high inflow rate is needed, ($\tau _{inf}$,$M_0$)=(5 Gyr,$10^9\:
M_{\odot }$).

\begin{figure}
\epsfxsize=\linewidth
\epsfbox{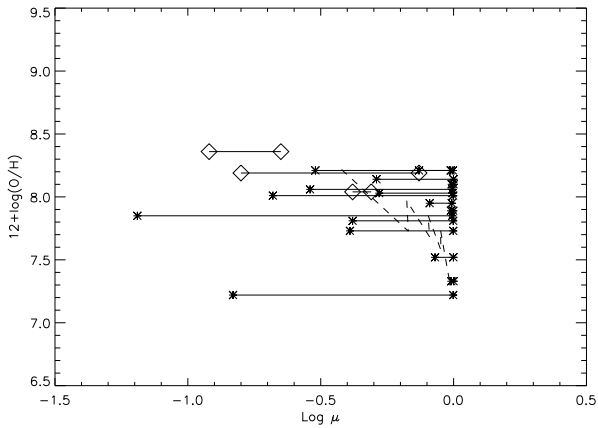}
\caption{Another model using yield set 2a, employing burst masses equal to 
$8\times 10^6\: M_{\odot }$, $M_0=10^9\: M_{\odot }$, $\tau _{inf}$=5 Gyr and 
$W_{ISM}$=8. Note the 'sawtooth' behavior caused by the high inflow rate.} 
\label{gaslargescatter}
\end{figure}

\begin{figure}
\epsfxsize=\linewidth
\epsfbox{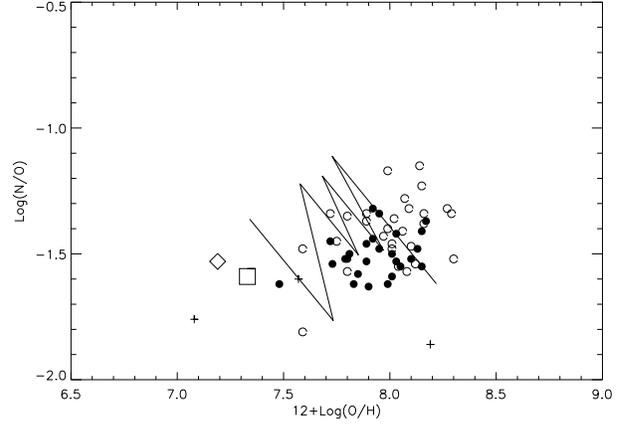}
\caption{The evolution of N/O-O/H for the same model as in fig.\ 
\ref{gaslargescatter}.} \label{NOlargescatter}
 \end{figure}
 
The results are shown in figs.\ \ref{gaslargescatter} and
\ref{NOlargescatter}. Note that the high outflow/inflow rate forces N/O to
increase dramatically with increasing O/H, because inflow
of primordial gas decreases O/H, but not N/O. At the same time, the strong
outflow decreases the O abundance more than the N abundance, since the
composition of the ISM just before each second pair-burst is O enhanced,
hence giving an O enhanced wind (though defined and treated as an ordinary
wind).
 
The corresponding gas fraction plot fits NGC 6822 and SMC, but the LMC gas 
fraction and abundance are never
reached. Note the 'sawtooth' behavior  in the gas fraction plot. During the long 
interburst period, the high inflow rate deposits a large amount of H in the
ISM, but no O, hence decreasing O/H until the next burst enriches the ISM 
again. However, the gas fraction does not increase, because the inflow is 
balanced by the high outflow rate.
From these last results, it seems to be possible to explain both the scatter
in N/O and the gas fractions simultaneously, but note that in our struggle to
explain the gas fractions of the three selected galaxies, some 3-4 other 
dwarf galaxies are not fitted within the intervals. Note also that the closed 
model explained those gas fractions quite well!

The results using yield set 3 are displayed in fig.\ \ref{infpad}. 
The gas fractions of the three well-known galaxies are properly fitted, 
though the same remark as above has
to be made, namely that the fits are outside the intervals of 3-5 other
dwarf galaxies.

\begin{figure}
\epsfxsize=5cm
\epsfbox{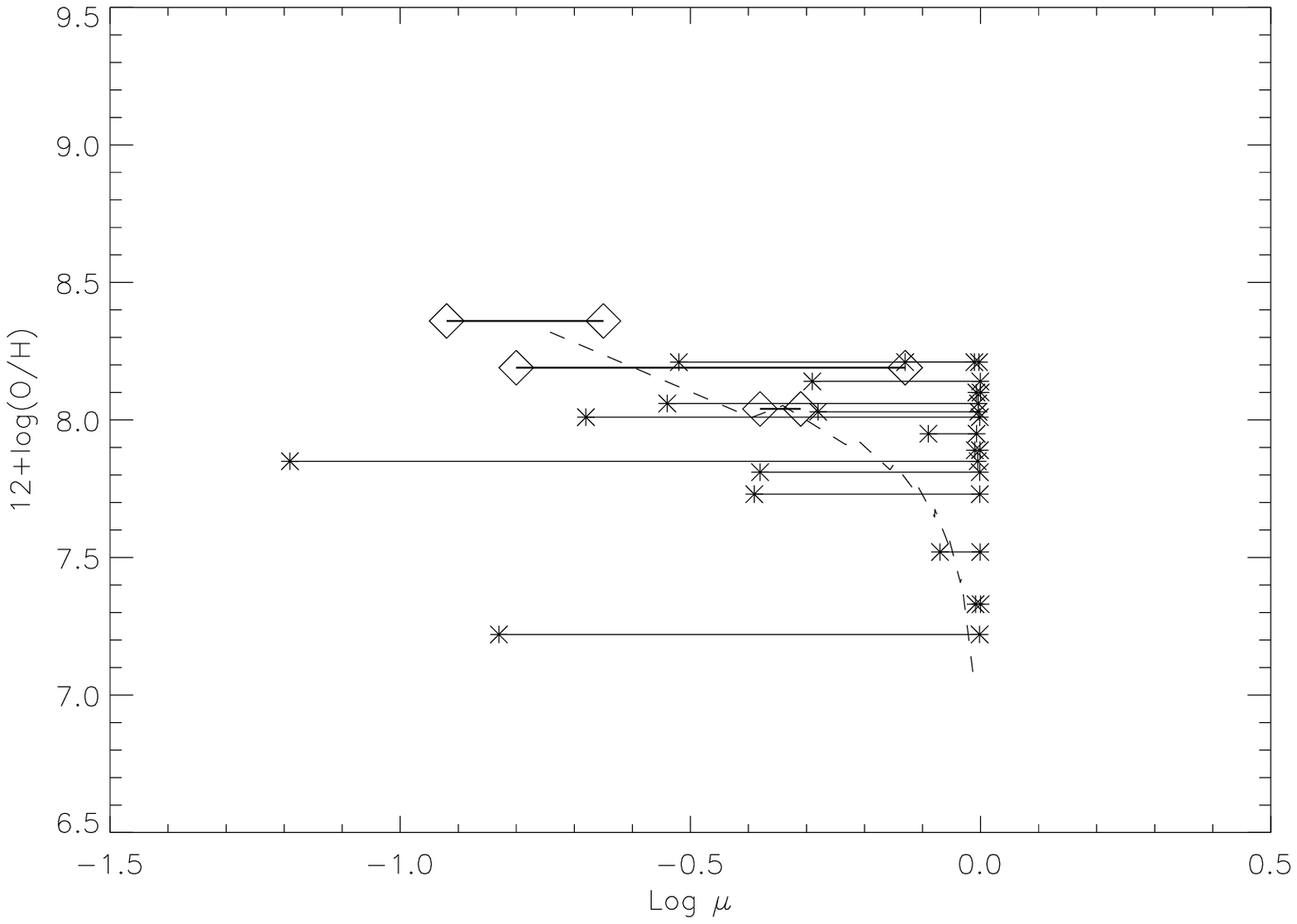} \vfill
\epsfxsize=5cm
\epsfbox{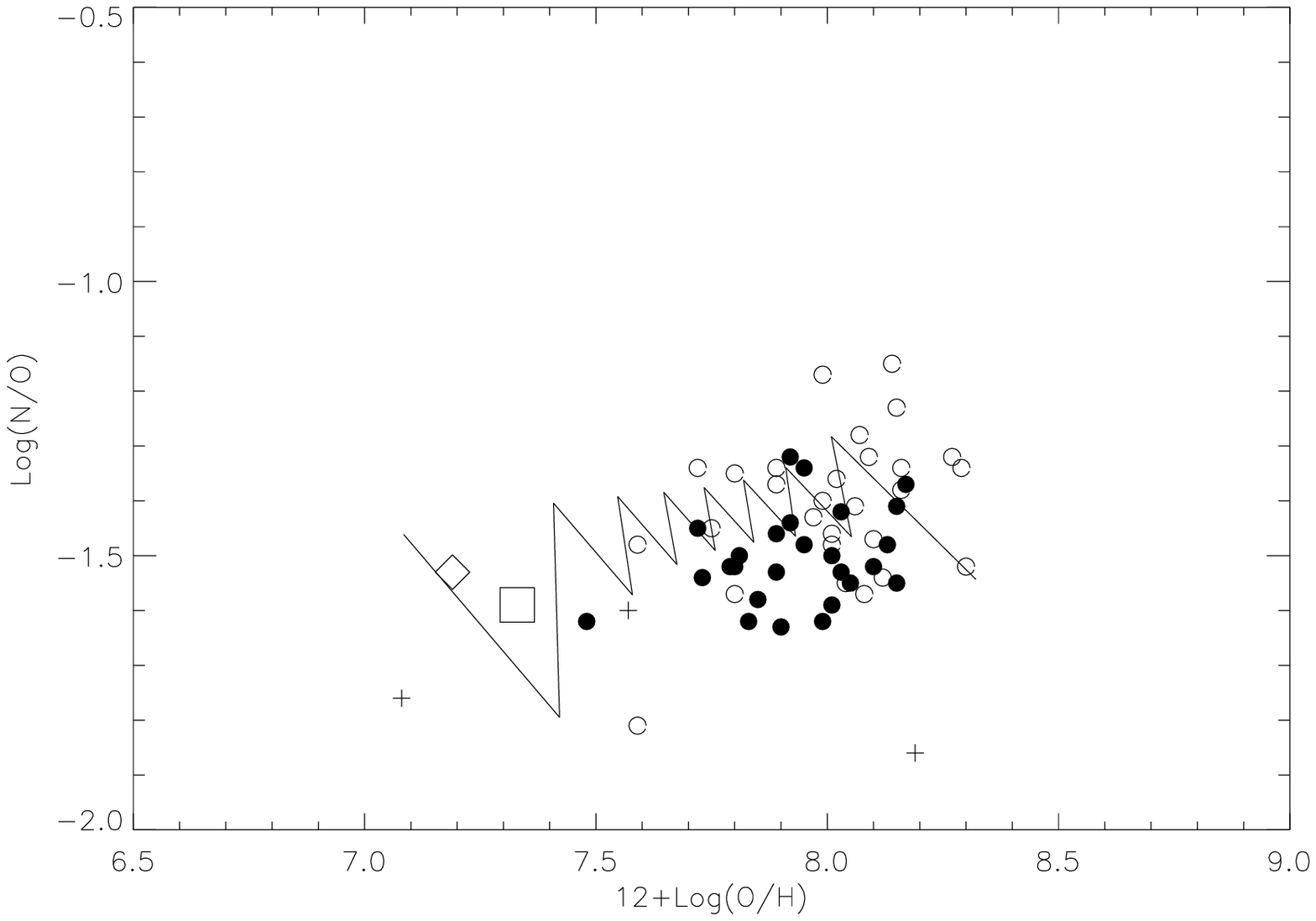} \vfill
\epsfxsize=5cm
\epsfbox{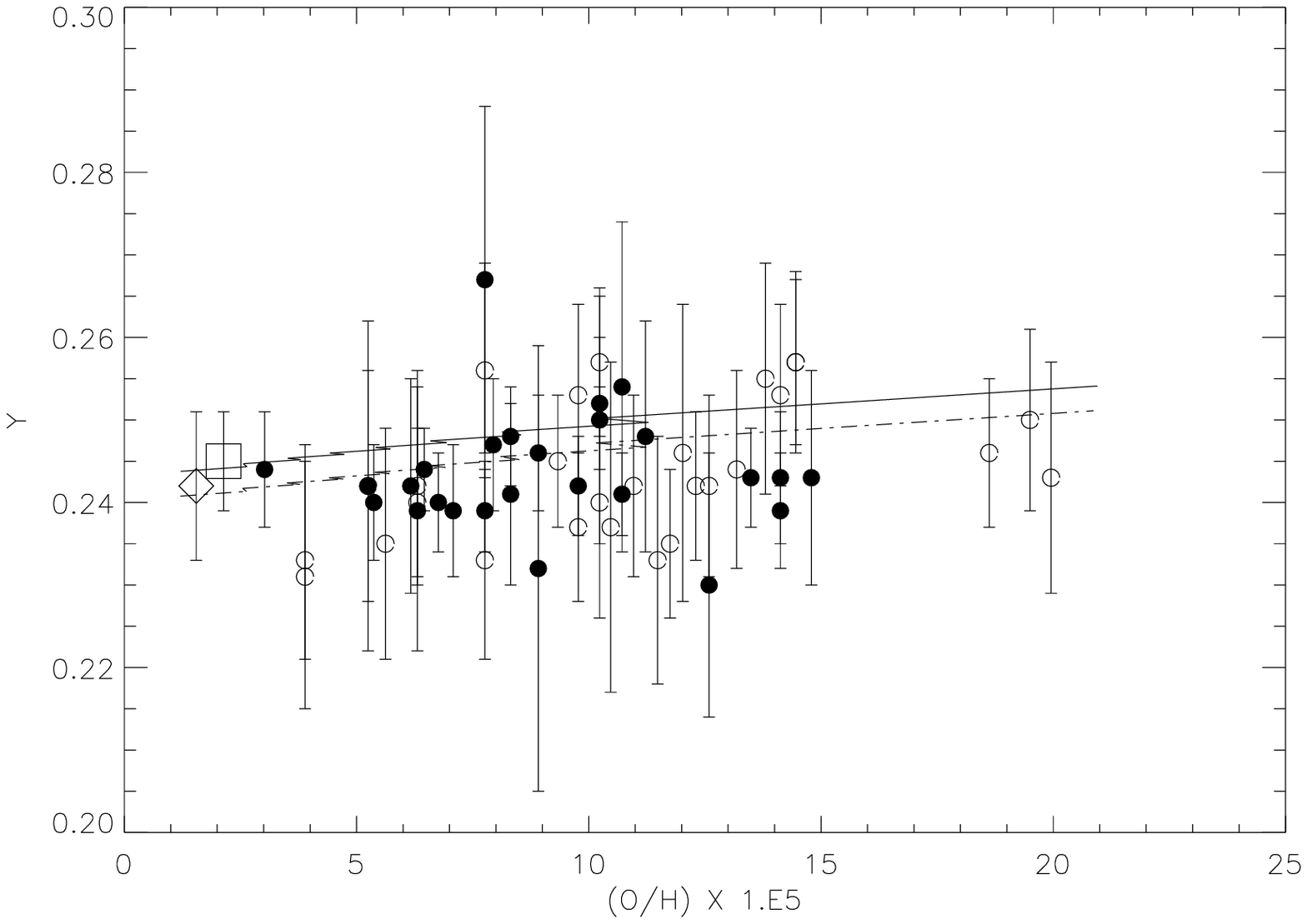} 
\caption{Results using yield set 3. For the model represented here
the adopted burst masses are $6\times 10^6\: M_{\odot }$ and
$M_0=7\times 10^8\: M_{\odot }$. Finally, $W_{ISM}$=5 and 
$\tau _{inf}$=5 Gyr.} \label{infpad}
\end{figure}

The right level of N/O is obtained. Note in particular the increasing N/O in
the left plot, following the trend of the observations. As for set 2a the
scatter in N/O is rather small, and insufficient to explain the observed
scatter. 

The four Y-O/H plots have not yet been discussed, but it may be done in a few 
words. The model predictions match the observations quite well, but they
are of course still showing the difference in slope between set 2a (or 2b) and 3.
\section{Discussion} \label{conc}
Numerical models calculating the chemical evolution of gas-rich dwarf galaxies
have been presented. The models have been fitted to a sample of abundance and gas 
fraction observations.  A chemical evolutionary model has to fit these 
observations simultaneously.
It is clear that a realistic model is not just a couple of equations including
a lot of parameters that one can change until all data are fitted. One has to
remember that the equations are applied to a physical system, obeying physical
laws. Hence, before doing any calculations, some considerations were made 
about ejecta dispersal and mixing processes. It was found that the processes 
involved are complicated and no complete theory exists yet. However, the four 
most important features in the enrichment process are found to be stellar 
ejecta, giant \hbox{H\,{\sc ii}}-regions, wind-driven superbubbles and 
SN-driven supershells. Wind-driven superbubbles arise shortly after a burst due 
to the strong stellar winds of massive stars. The radii of these superbubbles 
are in general smaller than the radii of the \hbox{H\,{\sc ii}}-regions. As 
the massive stars explode as SNe, a supershell is swept-up, soon catching up 
on the superbubble. However, a simple calculation showed that their radii 
become comparable to the radii of \hbox{H\,{\sc ii}}-regions only in the late 
stages of \hbox{H\,{\sc ii}}-region existence. Hence, it is suggested
that observed emission lines, used for abundance determinations, arise in the
ionized medium outside of the superbubble/supershell. Numerical
hydrodynamical models indicate that SN-ejecta always stay within the supershell.
If this is true, the observed abundances are not affected by the SN-ejecta from the 
stars producing the  \hbox{H\,{\sc ii}}-region,
hence being typical for the ISM.

The details of supershell evolution are still not known, but both theory and 
observations support star formation in expanding shells. From theory, it is 
expected to start not earlier than about 20 Myr after the starburst, resulting in 
star formation involving a mass comparable to the mass of the burst that 
initiated the supershell, hence appearing as a double burst. 
The double bursting mode of our numerical model ensures the appearance of 
scatter in the N/O-O/H plane, according to the time-delay idea, 
namely that N is released some time after O. The time 
interval between the two bursts of a pair is tuned to give maximum scatter. It 
is found that this timescale is comparable to the timescale of star formation 
in an expanding supershell. The requirement for the 'shell burst' to be 
the second pair-burst is that the O-rich ejecta mix into the supershell. 
Because of the poorly understood physics of supershells
and mixing processes, this assumption should be seen so far as a working 
hypothesis. 
All bursts are assumed to be instantaneous, hence representing short but intense
star formation events.  
The closed model is able to explain the observations of N/O-O/H in
both scatter and level. If assuming the upper limit of hot bottom burning to
be 5$M_{\odot }$, $\alpha $=1.1 is used to explain the observations, where 
$\alpha $ is the mixing length parameter. This is a rather low value compared 
to the results of recent
works however \cite{van_den_hoek:1997,marigo:1998}, favouring a value 
close to 2.

Using yield set 3 (the Padova set), it is necessary to extrapolate the
primary N yields below Z=0.008 to obtain a higher N yield. Otherwise, the 
level of N/O becomes too low compared to the observations. The 
primary N production has to be increasing with decreasing metallicity.
Calculations of stellar yields for metallicities lower than Z=0.008 are 
definitely desired to quantify this..

A very important conclusion is that no primary N production in massive stars
is needed to explain the observations. Intermediate mass stars are in position
to produce a sufficient amount of primary N. 

For the Y-O/H observations, a linear trend is visible,
and a linear fit gives $dY/dZ=2.63\pm 2.21$ and $Y_p=0.238\pm 0.004$.
These values are consistent with those of Izotov et al. \shortcite{izotov:1997} 
within the uncertainties. The closed model is found to explain the Y-O/H 
observations perfectly, only with different slopes, depending on the 
yield set in use. In all cases the slopes are within the uncertainties. As seen by 
inspection of the true He yields for the three
sets in table\ \ref{totalyield}, one finds the explanation for the slope
difference to be that the He yields of set 3 are 2-3 times higher than those
of set 2a or 2b. 

It is noteworthy that the Y-O/H relation was fitted using exactly
the same parameters as for the N/O-O/H fitting. 
The problem arises when fitting O/H-$\mu $ data. It is argued that 
the observed gas fractions are actually lower limits, because dark matter is
implicitly included in dynamical mass estimates and molecular hydrogen ignored. 
Hence, it is found useful
to calculate upper limits using M/L estimates from starburst evolutionary 
models. For most objects, very extreme upper limits have to be used, assuming
the galaxies to experience maximum luminosity of their first burst, except
for three galaxies where the known star formation histories allow us to adopt 
more moderate and realistic upper limits. If star formation histories are found for 
a larger 
sample of dIrrs and even BCGs constraining chemical evolution models with
observed gas fraction intervals may eventually turn out to be extremely useful. 
The closed models do not reproduce the gas fractions of the three well-known 
objects, 
even if a lower IMF-cutoff, equal to 0.01 $M_{\odot }$ instead of 0.1 
$M_{\odot }$ is adopted. Hence, open models are considered, allowing gas to 
escape or to accrete on to
the galaxy. Two kind of winds, enriched and ordinary, have been used. The 
results when incorporating enriched winds are not in accordance with the 
observed level of N/O and the Y-O/H fitting is not satisfactory, when the gas 
fractions of the three well-known systems are fitted. Hence, it is concluded 
that our models employing enriched winds is in conflict with the observations.

The next step is to include ordinary instead of enriched winds. Both N/O-O/H 
and Y-O/H are fitted, with results resembling those of the closed model. 
Unfortunately, only the gas fraction of one of the three selected galaxies is 
fitted, not differing much from the results of the closed model. 
To check the inclusion of ordinary winds, the outcome of the model is compared
to the results of a simple analytical model, employing continuous star 
formation. A close resemblance is found between the two models at low 
metallicities, but at higher metallicities, the numerical model seems to have 
problems in getting the yield down. The difference may be caused by the 
behavior of starbursts at a low absolute gas 
mass. This is confirmed when comparing the numerical model, changed slightly
to employ continuous star formation, to the analytical model, displaying almost 
identical outputs. Thus, it is concluded that it is important to specify 
clearly whether instantaneous bursts or continuous star formation is used, when 
including ordinary winds. Instantaneous bursts resemble the 
intense bursts of BCGs, whereas the more moderate bursts of dIrrs are 
better explained using a continuous SFR.

Finally, inflow and ordinary winds were included. The results are in accordance
with all observations, except that it is difficult to obtain the right gas 
fractions and N/O scatter simultaneously. Only for quite extreme parameter 
choices as in fig.\ \ref{gaslargescatter} and \ref{NOlargescatter}, one may be 
succesful. It may be important to note that the results, when 
using the combined inflow/ordinary wind model, show the upturn in N/O to be 
more pronounced, than it was for the closed model. 

One question is unavoidable: which model is preferred? It is impossible to give 
an unambiguous answer. Dwarf galaxies are different, both in mass and 
appearance. Some are explained well by a closed model, others need a 
combination of ordinary winds and inflow. This is true for both dIrrs an 
BCGs. However, for dIrrs one should prefer to adopt a more continuous SFR
before fitting the observations.

In all cases, the model including enriched winds seems to be ruled out, since it 
is in direct conflict with the observations, as also found by
e.g. Carigi et al. \shortcite{carigi:1998}.

\end{document}